\documentclass[journal, ,x11names]{IEEEtran}
\usepackage{setspace}

\usepackage{subfigure}
\usepackage{amsmath,amsfonts}
\usepackage{algorithmic}
\usepackage{algorithm}
\usepackage{array}
\usepackage{subfig}
\usepackage{textcomp}
\usepackage{stfloats}
\usepackage{url}
\usepackage{verbatim}
\usepackage{graphicx}
\usepackage{cite}
\usepackage{subfloat}
\usepackage{tikz}
\usepackage{bm}

\usepackage{color,xcolor}
\usepackage{caption}
\usepackage{pgfplots}
\usetikzlibrary{calc,arrows.meta,positioning}
\usetikzlibrary{shapes.arrows}
\usetikzlibrary{fit,backgrounds} 
\usepackage{standalone}
\usepackage[numbers,sort&compress]{natbib}
\usepackage{setspace}

\begin{document}

\title{Robust Image Semantic Coding with Learnable CSI Fusion Masking over MIMO Fading Channels}

\author{Bingyan Xie, Yongpeng Wu,~\IEEEmembership{Senior Member,~IEEE,} Yuxuan Shi, Wenjun Zhang,~\IEEEmembership{Fellow,~IEEE}, \\Shuguang Cui,~\IEEEmembership{Fellow,~IEEE}, and Merouane Debbah,~\IEEEmembership{Fellow,~IEEE}
        % <-this % stops a space

\thanks{(Corresponding author: Yongpeng Wu.)}
\thanks{Bingyan Xie, Yongpeng Wu, and Wenjun Zhang are with the Department of Electronic Engineering, Shanghai Jiao Tong University, Shanghai 200240, China (e-mail:bingyanxie, yongpeng.wu, zhangwenjun@sjtu.edu.cn).}% <-this % stops a space
\thanks{Yuxuan Shi is with the School of Cyber and Engineering, Shanghai Jiao Tong University, Shanghai 200240, China(e-mail:ge49fuy@sjtu.edu.cn).}
\thanks{Shuguang Cui is with the School of Science and Engineering, the Future Network of Intelligence Institute, and the Guangdong Provincial Key Laboratory of Future Networks of Intelligence, the Chinese University of Hong Kong (Shenzhen), Shenzhen 518172, China, and with Shenzhen Research Institute of Big Data, Shenzhen 518172, China, and with Peng Cheng Laboratory, Shenzhen 518066, China (e-mail: shuguangcui@cuhk.edu.cn).}
\thanks{M. Debbah is with KU 6G Research Center, Khalifa University of Science and Technology, P O Box 127788, Abu Dhabi, UAE (email: merouane.debbah@ku.ac.ae) and also with CentraleSupelec, University Paris-Saclay, 91192 Gif-sur-Yvette, France.}
}

%{Shell \MakeLowercase{\textit{et al.}}: A Sample Article Using IEEEtran.cls for %IEEE Journals}

%\IEEEpubid{IEEE}

\maketitle

\begin{abstract}
Though achieving marvelous progress in various scenarios, existing semantic communication frameworks mainly consider single-input single-output Gaussian channels or Rayleigh fading channels, neglecting the widely-used multiple-input multiple-output (MIMO) channels, which hinders the application into practical systems. One common solution to combat MIMO fading is to utilize feedback MIMO channel state information (CSI). In this paper, we incorporate MIMO CSI into system designs from a new perspective and propose the learnable CSI fusion semantic communication (LCFSC) framework, where CSI is treated as side information by the semantic extractor to enhance the semantic coding. To avoid feature fusion due to abrupt combination of CSI with features, we present a non-invasive CSI fusion multi-head attention module inside the Swin Transformer. With the learned attention masking map determined by both source and channel states, more robust attention distribution could be generated. Furthermore, the percentage of mask elements could be flexibly adjusted by the learnable mask ratio, which is produced based on the conditional variational interference in an unsupervised manner. In this way, CSI-aware semantic coding is achieved through learnable CSI fusion masking. Experiment results testify the superiority of LCFSC over traditional schemes and state-of-the-art Swin Transformer-based semantic communication frameworks in MIMO fading channels.

\end{abstract}


\section{Introduction}\label{s1}
\IEEEPARstart{A} new communication paradigm, called semantic communication, has emerged recently. Being regarded as one of the promising technologies in the sixth-generation (6G) wireless communications, semantic communication has great potential in many practical scenarios, e.g. intelligent humanoid robots, Internet of Vehicles, and cloud computing, etc. Unlike the traditional schemes, such advanced paradigm is a bold attempt to deeply integrate artificial intelligence (AI) into the wireless communication system designs \cite{6G,sc, qin2021semantic}. Instead of merely focusing on the accurate bit recovery, semantic communication incorporates the compact inner semantic representation, which greatly helps the system save the transmission overhead \cite{zhang2022toward}.

In the late 1940s, Shannon \cite{Shannon} has proposed the three-level communication framework. The second level, which is named as semantic communication, has been explored theoretically over the past seven decades \cite{Carnap1952,background,mimo}. Nevertheless, these investigations were mainly based on texts and logistic probability measure, which were restricted for nowadays multi-modal media. At present, deep learning (DL) methods have been employed into the semantic communication system designs for its potential in understanding the human world \cite{error,robust,DVST,flsc,opti}. Among these, DL-based semantic communications with joint source and channel coding (JSCC) serve as a promising solution to reduce transmission redundancy, especially for images and videos. For example, Hu et al. have designed a masked vector quantized-variational autoencoder and constructed a discrete shared codebook for encoded image semantic representation. Wang et al. have proposed a deep video semantic transmission structure based on the nonlinear transform and conditional coding to adaptively extract semantic features according to the source entropy.

\begin{figure}[htbp]
	\centering
	\includegraphics[width=3.5in]{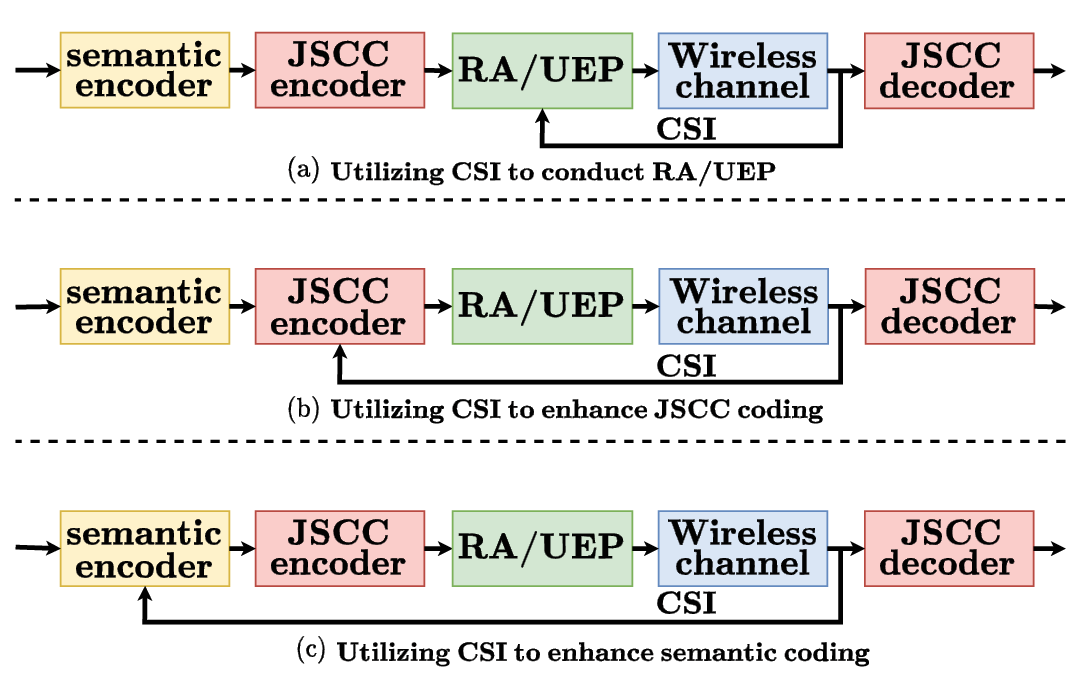}
	\caption{Different usages of feedback CSI. (a) and (b): Common semantic communication schemes with CSI feedback. (c): A novel scheme fusing CSI as side information into the semantic encoder.}
	\label{fig_a}
\end{figure}

Although JSCC-based semantic communication frameworks have made enormous progress in multi-modal media transmission, they mainly focus on the single-input single-output additive white Gaussian noise (AWGN) channels or stable Rayleigh fading channels, which impose great limitations in practical scenarios, e.g. multiple-input multiple-output (MIMO) fading channels. By considering the impact of MIMO communication systems in nowadays wireless transmission, channel state information (CSI) is one of the crucial factors. Therefore, with the aim to construct a MIMO system-based semantic communication framework, it inevitably brings the following two challenges. One is how to acquire accurate CSI through time-varying channel estimation \cite{chinese} and timely CSI feedback. Previous works \cite{CE, CSINET} have successfully applied deep neural networks to efficiently conduct channel estimation and CSI feedback, separately. The other is how to properly leverage CSI to combat fading and noise. To tackle this challenge, a variety of works have considered semantic communication frameworks with CSI in different perspectives. These frameworks typically encode the original information via a semantic encoder and a JSCC encoder, in which the former is for semantic extraction and the latter is for semantic compression and transmission. Existing works mainly focus on the interaction of CSI and transmitter, which is illustrated in Fig. \ref{fig_a}. Among them, some works \cite{SCAN, LIT} utilize the feedback CSI after the channel encoder, e.g. unequal error protection (UEP) and rate allocation (RA) modules, as shown in Fig. \ref{fig_a}(a). For instance, Yao et al. \cite{LIT} have introduced an extra learnable CSI token set which is directly embedded into the Transformer block for rate allocation and stream mapping. We note that it is actually a separated coding design since both encoders have no idea of the channel state. Meanwhile, in Fig. \ref{fig_a}(b), introducing MIMO CSI into the JSCC encoder is another common strategy for utilizing CSI in semantic communication. Jiang et al. \cite{SVC} have blended CSI into the JSCC encoding and quantization stage to ensure the keypoint protection for the latter incremental redundancy hybrid automatic repeat request. Wu et al. \cite{DJSCCF} have refined CSI as channel heatmap and incorporated it into the Vision Transformer (ViT)-based JSCC encoder \cite{Dosovitskiy} to automatically adapt to time-varying MIMO channel conditions. This combination though enables the codewords robust to the channel environments, to some extent, it still performs suboptimal since the extracted semantics are unaware of channel conditions. Therefore, a straightforward idea is to merge the feedback CSI as side information into the semantic encoder, as presented in Fig. \ref{fig_a}(c). This idea is inspired by previous works \cite{ACNN,TransFusion,Transid,Decouple} in computer vision (CV) and neural language processing (NLP) which integrate side information into feature extractors for different downstream tasks. For example, in the TransReID \cite{Transid}, authors employ camera ID messages through plugging in side information embedding into the ViT structure to mitigate feature bias caused by view variations for object re-identification. It hence motivates us to fuse CSI into semantic encoder for robust feature extraction over MIMO fading channels.

However, how to make MIMO CSI fusion pose positive effect on semantic coding is an intractable problem. The high-dimensional and complex-valued MIMO CSI matrix brings destruction to the architecture of raw features when implemented with simple fusion schemes. The rich contents in side information compound the feature embedding space, making it more difficult to extract semantics precisely. This phenomenon is referred to as ‘feature invasion’ \cite{nova}. Current works \cite{ADJSCC,PADC}, such as attention DL-based joint source-channel coding (ADJSCC), predictive and adaptive deep coding (PADC), mainly focus on blending the low-dimensional side information into a semantic extractor for performance gain over the original deep JSCC \cite{DJSCC}. Specifically, the ADJSCC conceives signal-to-noise ratio (SNR) as side information and directly concatenates it with extracted semantics of the encoder. The nonlinear transform source-channel coding (NTSCC) \cite{NTSCC} utilizes the auxiliary latent variables as side information for directly incorporating feature patches with rate allocation tokens in Transformer. Nevertheless, the above side information fusion schemes are not work for the MIMO CSI fusion, since CSI is no longer a scalar or vector, but a matrix reflecting the interactions among different antennas. The increasing dimension of side information leads to drastic performance fluctuating when being integrated into the features. Some intuitive combination schemes, e.g. fusing CSI with features by abrupt concatenation or summation, will damage the spatial structure through directly flattening CSI into sequences.

To solve the feature invasion during MIMO CSI fusion, we offer a sophisticated design to efficiently combine the CSI and features. In this paper, we investigate a learnable CSI fusion semantic communication (LCFSC) framework, integrating feedback CSI with the semantic coding to generate more robust semantic features against MIMO fading. Swin Transformer \cite{Swin}, which utilizes the attention mechanism to enhance long-term feature extraction ability among contexts and achieves the state-of-the-art performance in many CV tasks, is a novel network backbone adopted in several existing semantic communication frameworks \cite{NTSCC, WITT}. In the LCFSC, such state-of-the-art Swin Transformer structure is also employed as the backbone of the semantic codec. We explore the in-depth mechanism of the attention module inside the semantic encoder and propose a non-invasive CSI fusion multi-head self attention (NI-CFMA) module for adequately combining CSI with semantic features. Different from the classic multi-head attention (MHA), NI-CFMA completes the integration via generating an attention masking map, which masks the elements in attention weights adaptively according to semantic relevance and channel conditions. The percentage of mask elements is controlled by a hyperparameter, i.e. mask ratio. Furthermore, to timely acquire the suitable mask ratio in NI-CFMA, the recurrent condition generation is proposed as a preprocessing stage to produce proper mask ratio in an unsupervised manner before JSCC encoding. In this way, feature elements which have relatively low semantic importance and will experience serious channel impairment can be discarded, resulting in a considerable performance increase in such a MIMO fading system. The contributions of this paper can be summarized as follows:

\begin{enumerate}
\item{LCFSC Framework:}
we propose a novel framework which efficiently integrates the feedback CSI of general 5G MIMO environment into a semantic communication system, which is based on Swin Transformer backbone. The main idea of LCFSC is derived from fusing CSI as side information into the semantic extractor to ensure the robust semantic coding. For a common non-stationary 5G wireless environment \cite{channel}, it achieves performance gain over 2 dB compared to existing works.
\item{CSI Fusion Masking Semantic Extractor:}
a sophisticated design of CSI fusion is provided in LCFSC. The original MHA module in Swin Transformer is substituted by a proposed non-invasive CSI fusion multi-head attention module. It provides suitable attention masking map which is jointly determined by the computed semantic relevance and feedback CSI. In this way, UEP inside the semantic coding can be realized through masking the unimportant and heavily disturbed semantic elements to generate more robust attention distribution.
\item{Learnable Mask Ratio:}
we further design a preprocessing stage to adaptively control the mask percentage of semantic attention elements in NI-CFMA by producing suitable mask ratio before JSCC encoding. A predefined stable mask ratio acts as the condition to help the unsupervised variational encoder generate the suitable mask ratio for the current state. In this way, the learnable mask ratio provides flexiblity for the NI-CFMA to produce robust semantic codewords against varying source and channel conditions. 
\item{Noise Purified Channel Estimation:}
we propose a noise purified channel estimator to produce accurate feedback CSI. It utilizes the coarse CSI through least square (LS) estimation as input and then purifies such coarse CSI into relatively fine CSI based on the noise space projection. A two-step training strategy is adopted to further ensure the accurate CSI acquisition in LCFSC.
\end{enumerate}

\begin{figure*}[htbp]
	\centering
	\includegraphics[width=6.5in]{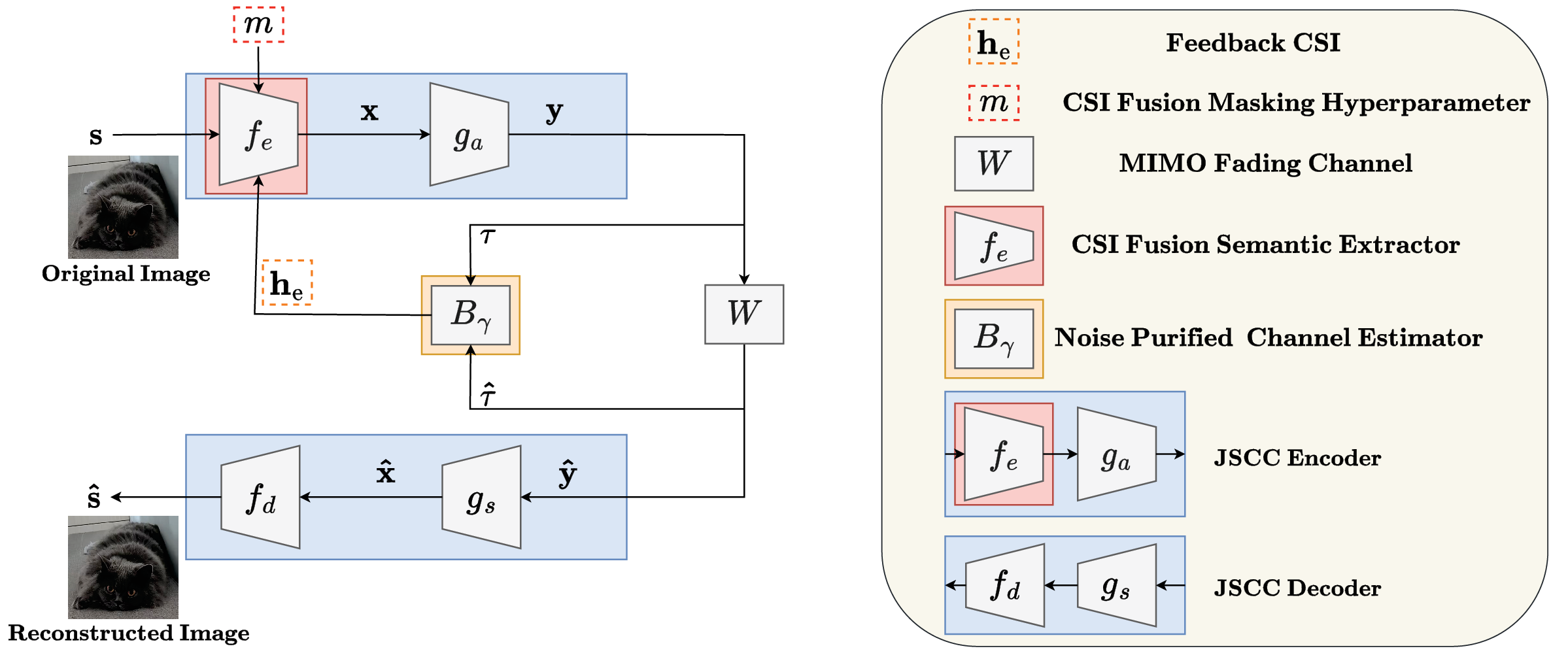}
	\caption{CSI fusion semantic communication (CFSC) framework.}
	\label{fig_1}
\end{figure*}

The rest of this paper is organized as follows. Section II introduces the system model and the framework of the CSI fusion semantic communication. Section III illustrates the detailed framework structures, including CSI fusion masking semantic extractor, noise purified channel estimator and other JSCC network modules. Section IV describes the LCFSC based on aforementioned JSCC-based structures with the modified recurrent condition generation stage. The corresponding training loss and strategies are also provided in this part. Section V demonstrates the superiority of the proposed networks through a series of experiments. Section VI concludes the paper.

Notational Conventions: $\mathbb{R}$ and $\mathbb{C}$ refer to the real and complex number sets, respectively. $\mathcal{N}\left (\mu, \sigma^2 \right)$ denotes a Gaussian distribution with mean $\mu$ and variance $\sigma^2$. $\mathbb{E}$ refers to the mathematical expectation. $\odot$ represents element-wise multiplication. Finally, $\left(\cdot\right)^{T}$ denotes the matrix transpose while $\left(\cdot\right)^{-1}$ is the matrix inverse.

\section{System Model and Proposed Framework}
Under a classic image wireless transmission scenario considering MIMO fading channels, we propose a CSI fusion semantic communication (CFSC) framework. It regards the feedback CSI as side information for the semantic encoder to enhance the robust semantic extraction against fading and noise for the latter transmission. In this section, we first provide the system model of wireless image transmission over MIMO fading channels. Then, the overall framework of CFSC is presented.
\subsection{System Model}
We consider a typical wireless image transmission problem, which aims at image reconstruction. Given a set with $N$ different images $\mathcal{S}=\left \{s_1,s_2,\cdots,s_N\right\}$, where each image $\mathbf{s}_i\in\mathbb{R}^{H\times W\times 3}$. The encoder at the transmitting end encodes the image set $\mathcal{S}$ into a codeword sequence set $\mathcal{Y}=\left \{y_1,y_2,\cdots,y_N\right\}$, where $\mathbf{y}_i\in \mathbb{R}^{C_{\mathrm{L}}}$ is the transmitted codewords of the $i$-th image with length $C_{\mathrm{L}}$. In this way, channel bandwidth ratio (CBR) is defined as $R=\frac{C_\mathrm{L}}{H\times W\times 3}$. After that, the codewords pass through the MIMO fading channel, which can be formulated as
\begin{align}
	\hat{\mathbf{y}}=\mathbf{h}\mathbf{y}+\mathbf{n}, 
\end{align}
where $\mathbf{y}\in\mathbb{R}^{N_{\mathrm{T}}\times \frac{C_{\mathrm{L}}}{N_{\mathrm{T}}}}$ is the reshaped codewords for MIMO transmission, $\hat{\mathbf{y}}\in\mathbb{R}^{N_{\mathrm{R}}\times \frac{C_{\mathrm{L}}}{N_{\mathrm{T}}}}$ is the received codewords, $\mathbf{h}\in\mathbb{C}^{N_{\mathrm{R}}\times N_{\mathrm{T}}}$ is the practical MIMO channel state matrix while $\mathbf{n}\in\mathbb{C}^{N_{\mathrm{R}}\times \frac{C_{\mathrm{L}}}{N_{\mathrm{T}}}}$ is a complex Gaussian noise matrix with mean $0$ and variance $\sigma^2$ of each element.

Finally, the decoder part translates the transmitted codewords into a recontructed image set $\hat{\mathcal{S}}=\left \{\hat{s}_1,\hat{s}_2,\cdots,\hat{s}_N\right\}$.

\subsection{Proposed Framework of CFSC}
The proposed CFSC framework is shown in Fig. \ref{fig_1}. First, the semantic extractor, $f_{e}(\cdot, \cdot, \cdot): \mathbb{R}^{H\times W\times C}\times\mathbb{C}^{N_{\mathrm{R}}\times N_{\mathrm{T}}}\times{[0,1]} \mapsto \mathbb{R}^{C_{\mathrm{H}}}$, encodes the raw picture, $\mathbf{s}$, with the help of side information including estimated CSI, $\mathbf{h}_{\mathrm{e}}\in\mathbb{C}^{N_{\mathrm{R}}\times N_{\mathrm{T}}}$, and a hyper-parameter called semantic mask ratio, $m$, into the semantic features, $\mathbf{x}\in\mathbb{R}^{C_{\mathrm{H}}}$, where $C_{\mathrm{H}}$ refers to the sequence length of the extracted semantic information. Then, the extracted semantics, $\mathbf{x}$, pass through $g_{a}(\cdot): \mathbb{R}^{C_{\mathrm{H}}} \mapsto \mathbb{R}^{C_{\mathrm{L}}}$ to combat fading and noise. The transmitted codewords are defined as $\mathbf{y}$. The whole encoding process can be expressed as

\begin{align}
	\mathbf{s}\xrightarrow[]{f_{e}(\cdot, \mathbf{h}_{\mathrm{e}}, m)}\mathbf{x}\xrightarrow[]{g_{a}(\cdot)}\mathbf{y}.
\end{align}

Before passing through the MIMO fading channel, $\mathbf{y}$ is reshaped as $N_{\mathrm{T}}\times \frac{C_{\mathrm{L}}}{N_{\mathrm{T}}}$. Then the received codewords are given as $\hat{\mathbf{y}}\in\mathbb{C}^{N_{\mathrm{R}}\times \frac{C_{\mathrm{L}}}{N_{\mathrm{T}}}}$. Meanwhile, to acquire the accurate CSI, we transmit pieces of known pilots, $\bm{\tau}\in \mathbb{R}^{N_{\mathrm{T}}\times N_{\mathrm{T}}}$, as well as the semantic codewords transmission. Then the transmitted and received pilots are utilized for channel estimation. The estimated CSI matrix can be illustrated as
\begin{align}
	\mathbf{h}_{\mathrm{e}}=B_{\bm{\Upsilon}}(\bm{\tau}, \bm{\hat{\tau}}), 
\end{align}
where $\bm{\hat{\tau}}\in \mathbb{R}^{N_{\mathrm{R}}\times N_{\mathrm{T}}}$ represents the received pilots, $B_{\bm{\Upsilon}}(\cdot,\cdot)$ is the channel estimator with weights, $\bm{\Upsilon}$.

At the decoder end, the received semantics, $\hat{\mathbf{y}}$, are reshaped as $C_\mathrm{L}$ and converted to real-valued sequence. Then the codewords pass through the channel decoder, $g_{s}(\cdot):\mathbb{R}^{C_{\mathrm{L}}} \mapsto \mathbb{R}^{C_{\mathrm{H}}}$, and then the semantic decoder, $f_{d}(\cdot): \mathbb{R}^{C_{\mathrm{H}}} \mapsto \mathbb{R}^{H\times W\times C}$, to recover the original image. The reconstructed image can be received by
\begin{align}
	\hat{\mathbf{y}}\xrightarrow[]{g_{s}(\cdot)}\hat{\mathbf{x}} \xrightarrow[]{f_{d}(\cdot)}\hat{\mathbf{s}},
\end{align}
where $\hat{\mathbf{s}}\in\mathbb{R}^{H\times W\times C}$ and $\hat{\mathbf{x}}\in\mathbb{R}^{C_{\mathrm{H}}}$ represent the reconstructed image and recovered semantics, respectively.

\begin{figure*}[htbp]
	\centering  %图片全局居中
	\subfigure[]{
		\includegraphics[width=0.67\linewidth]{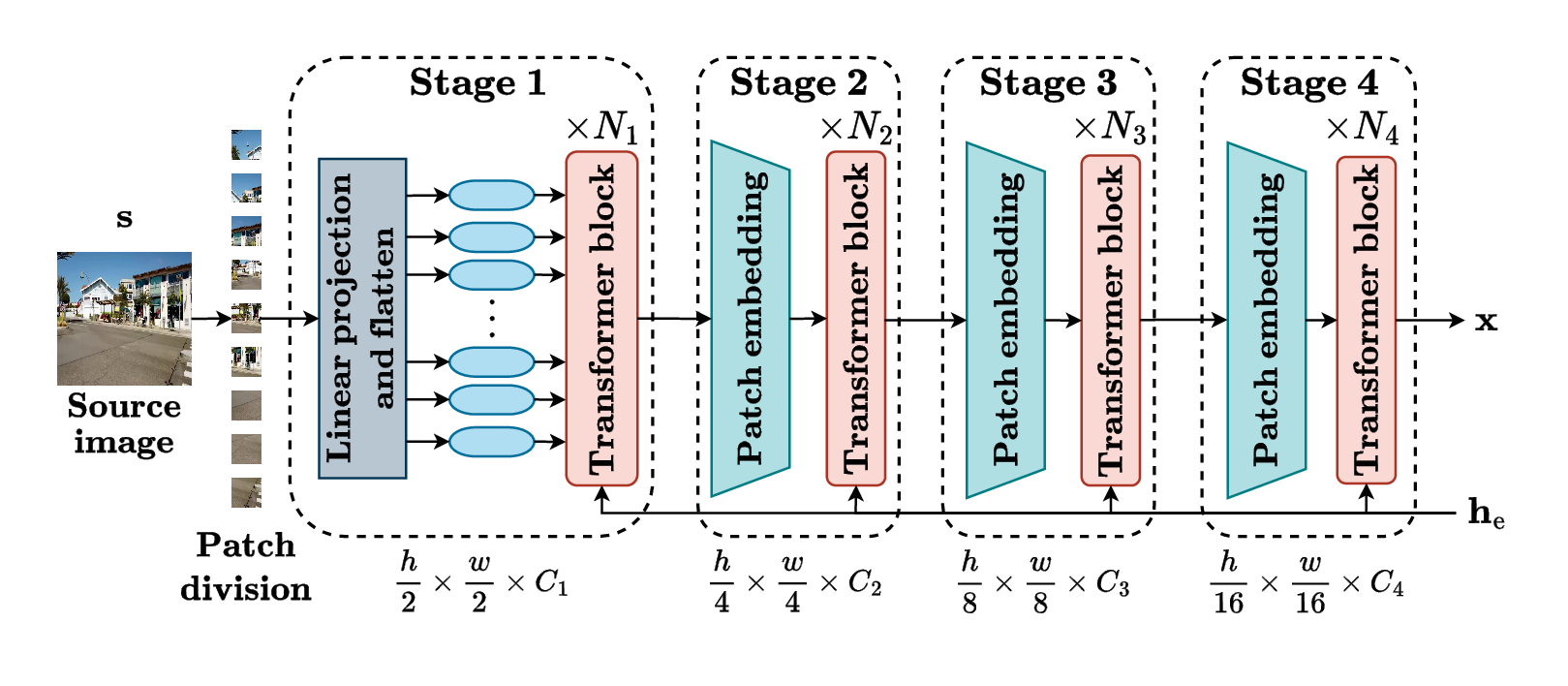}}
	\subfigure[]{
		\includegraphics[width=0.3\linewidth]{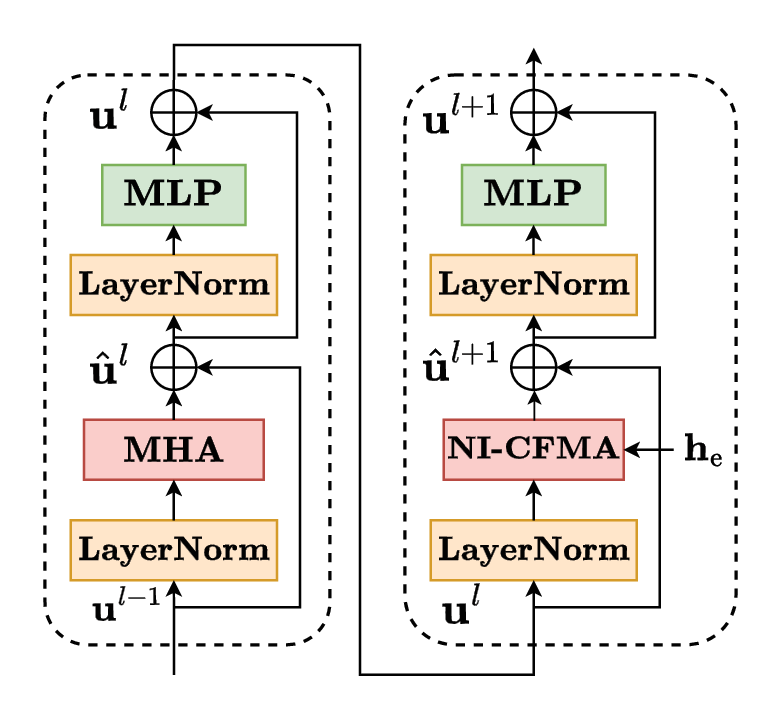}}
	\caption{(a) Network architecture of the CSI fusion semantic encoder $f_{e}$. (b) Two successive Swin Transformer blocks with different attention modules.}
	\label{fig_2}
\end{figure*}

\section{Detailed Structure of CFSC}
In this section, we give presentation to the components of CFSC in detail, including non-invasive CSI fusion semantic extractor, noise-basis channel estimator, and other JSCC network structures.

\subsection{Non-invasive CSI Fusion Semantic Extractor}

The proposed semantic extractor, $f_{e}$, is based on the Swin Transformer backbone \cite{Swin}. Fig. \ref{fig_2}(a) presents the structure of the semantic encoder, $f_{e}$. The source image, $\mathbf{s}$, is first splitted into $l_1=\frac{h}{2} \times \frac{w}{2}$ different image patches as the image sequence inputs. After the patch division, the sequence inputs are fed into a series of fully connected layers for linear projection and feature flattening. The outputs are feature embeddings with $c$ dimensions. Then, the feature embeddings pass through the hierarchical structure with several stages. Each stage $i$ is encapsulated by a patch embedding module and $N_i$ Transformer blocks which downsample the feature embeddings into $\frac{h}{2^i}\times \frac{w}{2^i}\times C_i$. Notably, the estimated CSI, $\mathbf{h}_\mathbf{e}$, is fed into the Transformer block as side information for semantic extraction enhancement. The structure of two successive Transformer blocks is shown in Fig. \ref{fig_2}(b). The two blocks are both constructed by the layer normalization layer, attention layer, and multi-layer perceptron (MLP) but have differences in the attention module. The second block substitutes the MHA with the NI-CFMA which conducts the sophisticated CSI fusion.

\begin{figure}[htbp]
	\centering
	\includegraphics[width=3.6in]{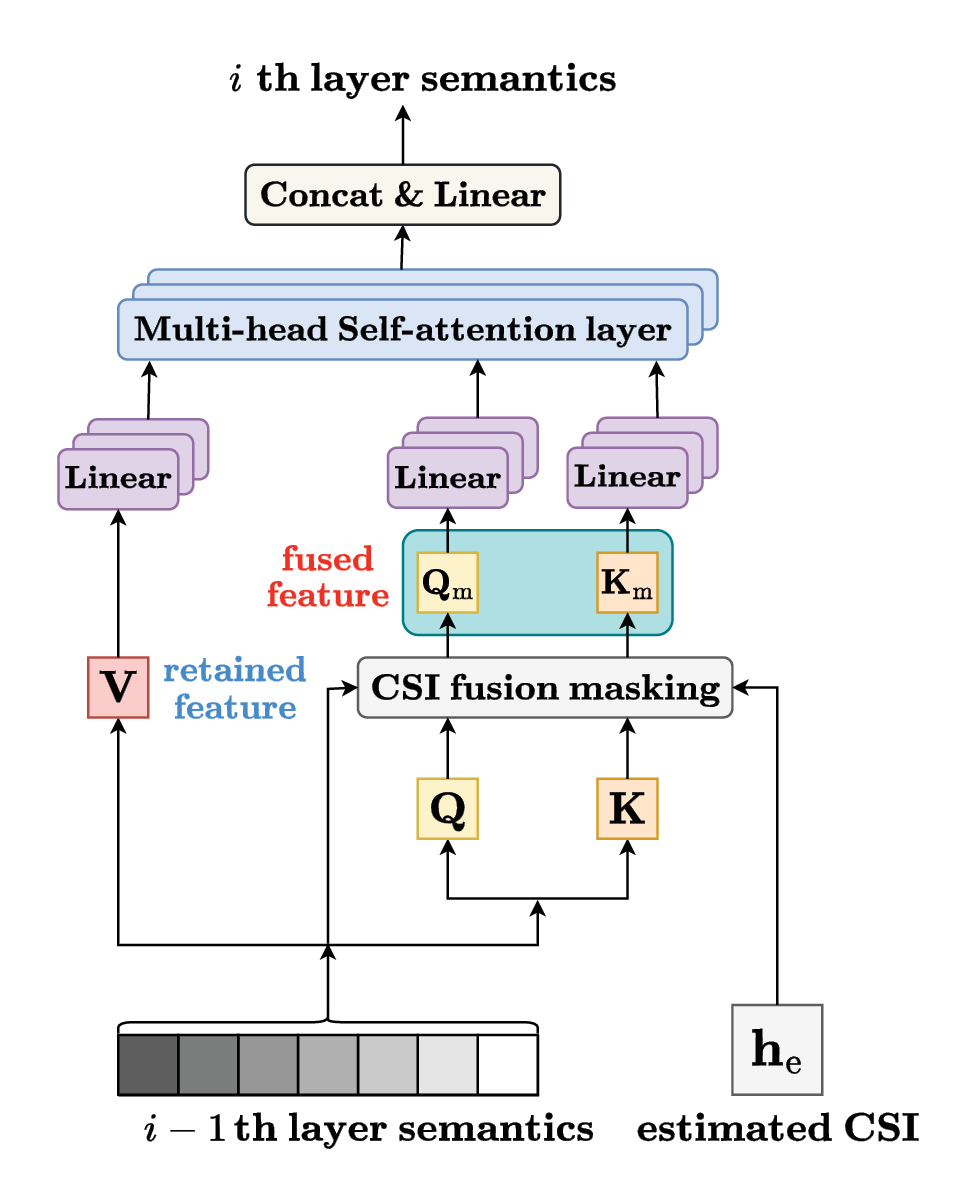}
	\caption{Non-invasive CSI fusion multi-head attention module.}
	\label{fig_3}
\end{figure}

The NI-CFMA adopts the structure of original MHA, which is an essential component embedded in the Transformer block. For completeness, we present the principle of the original MHA as followings. Through three separated linear projection networks, the input semantic sequence is transformed into three attention matrices, called query, $\mathbf{Q}\in \mathbb{R}^{D \times C_{\mathrm{H}}}$, key, $\mathbf{K}\in \mathbb{R}^{D \times C_{\mathrm{H}}}$, and value, $\mathbf{V}\in \mathbb{R}^{D \times C_{\mathrm{H}}}$, where $D$ refers to the number of patch embeddings. Matrix $\mathbf{V}$ reflects the importance of the input semantic sequence elements, while $\mathbf{Q}$ and $\mathbf{K}$ jointly represent the correlation of the arbitrary two input elements. Then, with the above three learnable matrices, the attention score, $\mathbf{A}\in \mathbb{R}^{D \times C_{\mathrm{H}}}$, can be obtained as the semantics extracted from the raw input, which can be illustrated as
\begin{align}
	\mathbf{A}(\mathbf{Q},\mathbf{K},\mathbf{V})=\varPhi(\frac{\mathbf{Q} \mathbf{K}^{T}}{\sqrt{d}}) \mathbf{V},
\end{align}
where $\varPhi(\cdot)$ represents the softmax function, $d$ refers to the scaling factor adjusting the attention weight, $\mathbf{Q} \mathbf{K}^{T}$. 

\begin{figure*}[htbp]
	\centering
	\includegraphics[width=7.0in]{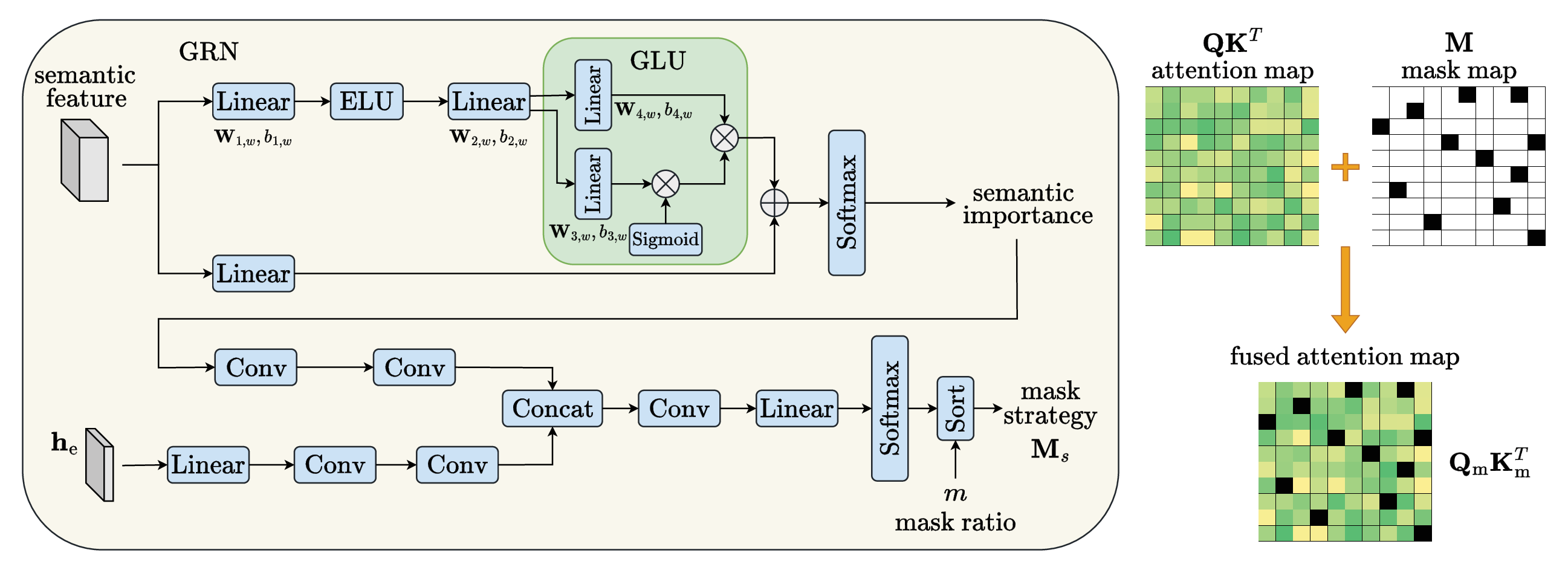}
	\caption{The structure of the CSI fusion masking model. Left is the overall network, while right illustrates the attention map masking strategy.}
	\label{fig_4}
\end{figure*}

The proposed NI-CFMA module is modified based on the MHA, which is shown in Fig. \ref{fig_4}. Jointly considering the accurate codewords transmission and effective semantic extraction, semantic elements with less importance and poor subchannel condition will inevitably impose negative effect on the semantic keypoints. This phenomenon triggers us to consider transferring the unequal protection into the semantic extraction, highlighting core semantic elements while supressing unimportant ones. We realize such protection via randomly masking the semantic elements, which sharpens the feature distribution with respect to semantics. This sharpness makes semantic distribution be more difficult to be affected than the original one \cite{Lee}. To properly fuse the feedback CSI into the attention module, we propose a CSI fusion masking model to polish the distribution of attention weights. To be specific, it generates the attention masking map according to the computed feature semantic importance along with the estimated CSI. A proper mask strategy benefits the feature robustness \cite{MAE} since only irrelevant elements are discarded, underlining the importance of remained semantics. Following this way, original feature embeddings $\mathbf{Q}$ and $\mathbf{K}$ are fused with the attention masking map, which we denote by $\mathbf{Q}_m$ and $\mathbf{K}_m$, respectively. The rest embedding $\mathbf{V}$ is purely computed by previous semantics and remains unchanged. Since the attention distribution is polished by CSI-aware masking strategy, which avoids the thorough distribution change caused by abruptly CSI fusion with features, without bringing extra difficulty for the latter semantic decoding.

More specifically, the structure of the CSI fusion masking model is shown in Fig. \ref{fig_4}, which can be divided into two steps. First, the semantic features pass through the gated residual network (GRN) \cite{TFT,GRN} and then obtain the semantic importance of each input feature patch, which reduces the input dimension and gives guidance to the later CSI fusion. Then, the refined semantic importance and the estimated CSI, $\mathbf{h}_{\mathrm{e}}$, are to generate the mask strategy. The GRN creates two paths: one is the direct path which is composed of a series of linear projection and activation layers. The other is the residual path which skips and connects with the main network path afterwards. Through the collaboration of two paths, it gets the importance level of input semantic features. Subsequently, two independent links composed of convolution neural networks map the semantic importance and $\mathbf{h}_{\mathrm{e}}$ into the same feature space. Then the two projected features are concatenated and pass through the later nonlinear layers. Finally, learned mask strategy, $\mathbf{M}_{\mathrm{s}}\in \mathbb{R}^{D\times D}$, is obtained through softmax and ascending sort process. The predefined mask ratio, $m=\frac{U}{D^2}$, is used to decide how much semantic elements will be masked randomly, where $U$ is the number of masked elements. With the mask strategy, $\mathbf{M}_{\mathrm{s}}$, we are able to mask the attention weights $\mathbf{Q}\mathbf{K}^T$ and compute the attention scores for the NI-CFMA. The actual mask matrix, $\mathbf{M}\in \mathbb{R}^{D\times D}$, can be expressed as
\begin{align}
	\mathbf{M}(i,j)= \begin{cases}
		0,  & \text{ if } \mathbf{M}_\mathrm{s}(i,j) <= m \times D\times D,\\
		1,  & \text{ if } \mathbf{M}_\mathrm{s}(i,j) > m\times D\times D,
	\end{cases}
\end{align}
where $\mathbf{M}(i,j)$ is the mask choice for the element in the $i$-th row and $j$-th column, $D\times D$ is the dimension of the attention weights, $\mathbf{Q} \mathbf{K}^T$.

Within $\mathbf{M}$ and $\mathbf{Q} \mathbf{K}^T$, the masked attention weight can be expressed as
\begin{align}
	\mathbf{Q}_{\mathrm{m}} \mathbf{K}_{\mathrm{m}}^T & = \mathbf{M}\odot(\mathbf{Q} \mathbf{K}^T).
\end{align}

The final computed attention scores for the NI-CFMA are shown as
\begin{align}
	\tilde{\mathbf{A}}(\mathbf{Q}_{\mathrm{m}},\mathbf{K}_{\mathrm{m}},\mathbf{V})=\varPhi(\frac{\mathbf{Q}_{\mathrm{m}} \mathbf{K}_{\mathrm{m}}^{T}}{\rho}) \mathbf{V},
\end{align}
where $\rho$ is the learnable substitute of the stable $\sqrt{d}$.

From the modified attention scores in Eq. (8), we find that the CSI fusion masking in the NI-CFMA focuses on producing better attention weight distribution aware of various channel states. Through the implementation of the NI-CFMA, intrinsic semantics are retained in $\mathbf{V}$ and elements with less semantic importance while faced with strong channel interference are to be masked in $\mathbf{Q}$ and $\mathbf{K}$ without inducing negative effects on other relatively more important semantics.

\subsection{Noise Purified Channel Estimator}
In order to acquire accurate feedback CSI in CFSC framework, noise purified network (NPN) is thus proposed to conduct channel estimation. As mentioned before, we utilize known pilots transmission along with the semantic codewords. At the first step, we jointly use the transmitted pilots $\bm{\tau}$ and received pilots $\bm{\hat{\tau}}$ to conduct the LS channel estimation, which can be written as
\begin{align}
	\hat{\bm{\tau}}&=\mathbf{h}\bm{\tau}+\mathbf{n}, 
\end{align}
\begin{align}
	\mathbf{h}_\mathrm{LS}&=\hat{\bm{\tau}}\bm{\tau}^{-1},
\end{align}
where $\mathbf{h}_\mathrm{LS}$ is the coarse CSI estimated by the LS method.

\begin{figure}[htbp]
	\centering
	\includegraphics[width=3.4in]{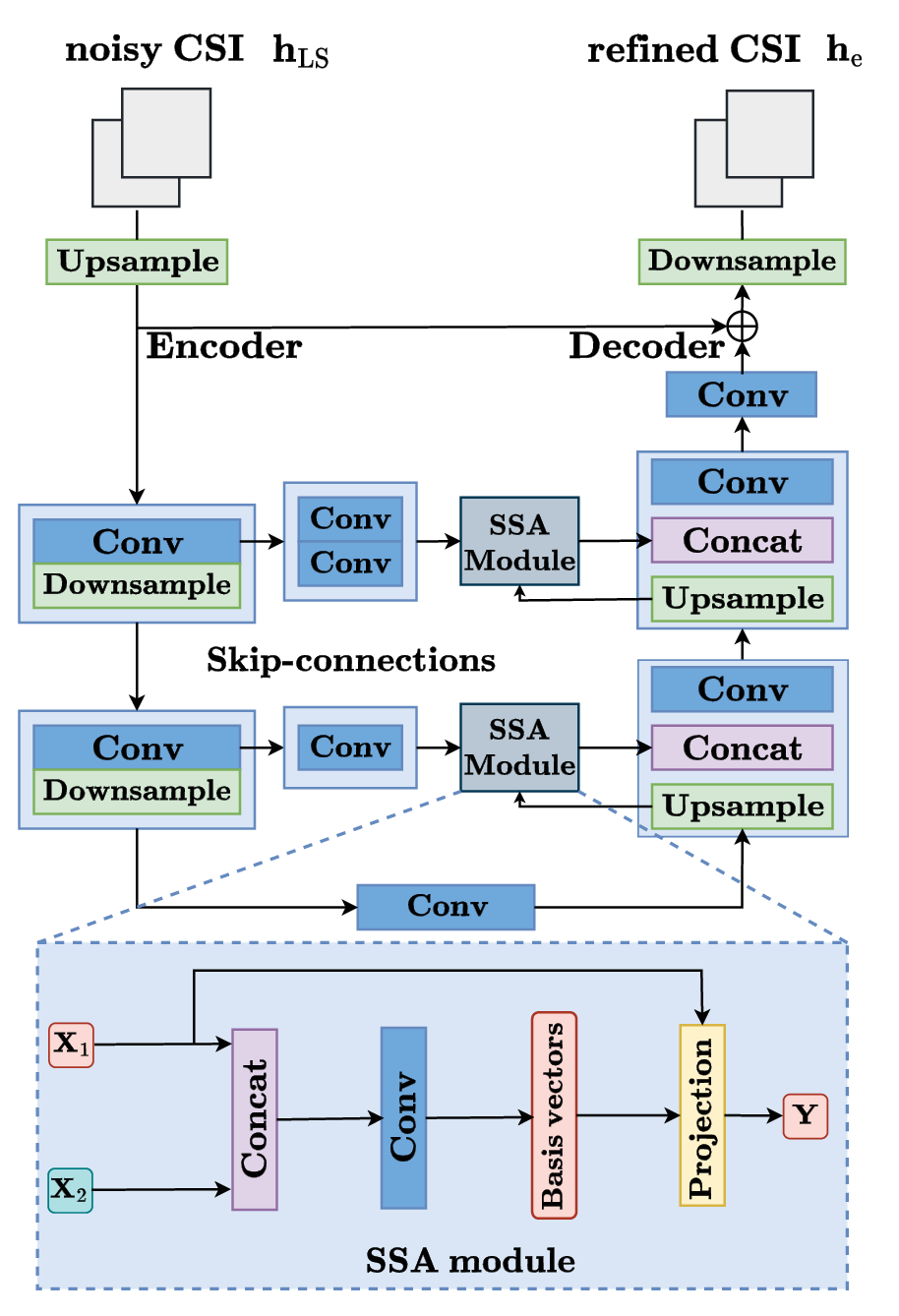}
	\caption{The architectures of the noise purified network.}
	\label{fig_5}
\end{figure}

In this way, the coarse estimated CSI can be obtained. After that, we feed $\mathbf{h}_\mathrm{LS}$ into NPN. It aims to produce the purified CSI, $\mathbf{h}_\mathrm{e}$, from the input CSI, which is similar to the image denoising. Noise purified networks \cite{NBNET} are originally employed for image denoising tasks in CV. Through generating subspace basis vector from the image feature maps and transforming feature maps into the signal space, the NPN achieves the efficient noise reduction performance for reconstructed images. To some extent, following the image denoising strategy, the coarse CSI can be polished as refined estimated CSI as well. The NPN is shown in Fig. \ref{fig_5}, which is based on the U-Net structure \cite{U-Net}. The encoder is escalated by several convolution layers and downsample layers while the decoder consists of several convolution layers, concatenate layers, and upsample layers, residually connected by the convolution layer and the subspace attention (SSA) module. The SSA enables the subspace noise basis generation and signal space projection. We denote $\mathbf{X}_1\in\mathbb{R}^{N_{\mathrm{R}}\times N_{\mathrm{T}}\times 2}$ and $\mathbf{X}_2\in\mathbb{R}^{N_{\mathrm{R}}\times N_{\mathrm{T}}\times 2}$ as the sampled features with the same dimension from the encoder part and decoder part, respectively. In the SSA, $N$ basis vectors, belonging to a $K$-dimensional signal subspace, $\mho\subset\mathbb{R}^N$, are estimated based on $\mathbf{X}_1$ and $\mathbf{X}_2$. To obtain the basis vectors, $\mathbf{V}_{\mathrm{b}}\in\mathbb{R}^{N\times K}$, we concatenate $\mathbf{X}_1$ and $\mathbf{X}_2$ along with the channel dimension as $\mathbf{X}$ first. Then the concatenated features pass through a residual convolutional network with $K$ output channels. The basis generation can be given as
\begin{align}
	\mathbf{V}_{\mathrm{b}} & = f_{\bm{\theta}}(\mathbf{X}_1,\mathbf{X}_2),
\end{align}
where $f_{\bm{\theta}}(\cdot)$ is the basis generation function with weights $\bm{\theta}$.

With the aforementioned matrix, $\mathbf{V}_{\mathrm{b}}$, through orthogonal linear projection, the noisy feature map $\mathbf{X}_1$ can be projected onto $\mho$.
\begin{align}
	\mathbf{Y} & = \mathbf{\psi}\mathbf{X}_1 = \mathbf{V}_{\mathrm{b}}(\mathbf{V}_{\mathrm{b}}^T\mathbf{V}_{\mathrm{b}})^{-1}\mathbf{V}_{\mathrm{b}}^T\mathbf{X}_1,
\end{align}
where $\mathbf{Y}\in\mathbb{R}^{N_{\mathrm{R}}\times N_{\mathrm{T}}\times 2}$ refers to the purified reconstructed features $\mathbf{X}_1$, $\mathbf{\psi}$ refers to the orthogonal projection matrix to the signal subspace, the normalization term $(\mathbf{V}_{\mathrm{b}}^T\mathbf{V}_{\mathrm{b}}) ^{-1}$ is essential to ensure the orthogonality between the arbitrary two elements of the basis vectors.

\subsection{Other Network Structure of the JSCC structure}

For the semantic decoder, $f_{d}$, it shares the same structure as the Swin Transformer decoder in \cite{Swin}. The JSCC components, $g_{a}$ and $g_{s}$, are also constructed by cascades of deep neural networks. In the CFSC, we also adapt the Channel ModNet \cite{WITT} pair. It enables the modulation and demodulation of the extracted and translated semantics while generating specific codewords for channel adaptation. 

\section{Learnable CSI fusion semantic communication framework}

In this section, to further provide flexibility for the NI-CFMA by generating suitable attention mask map aware of source and channel distribution, we let the mask ratio in CFSC framework learnable via a recurrent condition generation stage. The overall designs are collected under the name learnable CSI fusion semantic communication (LCFSC).

\begin{figure}[htbp]
	\centering
	\includegraphics[width=3.5in]{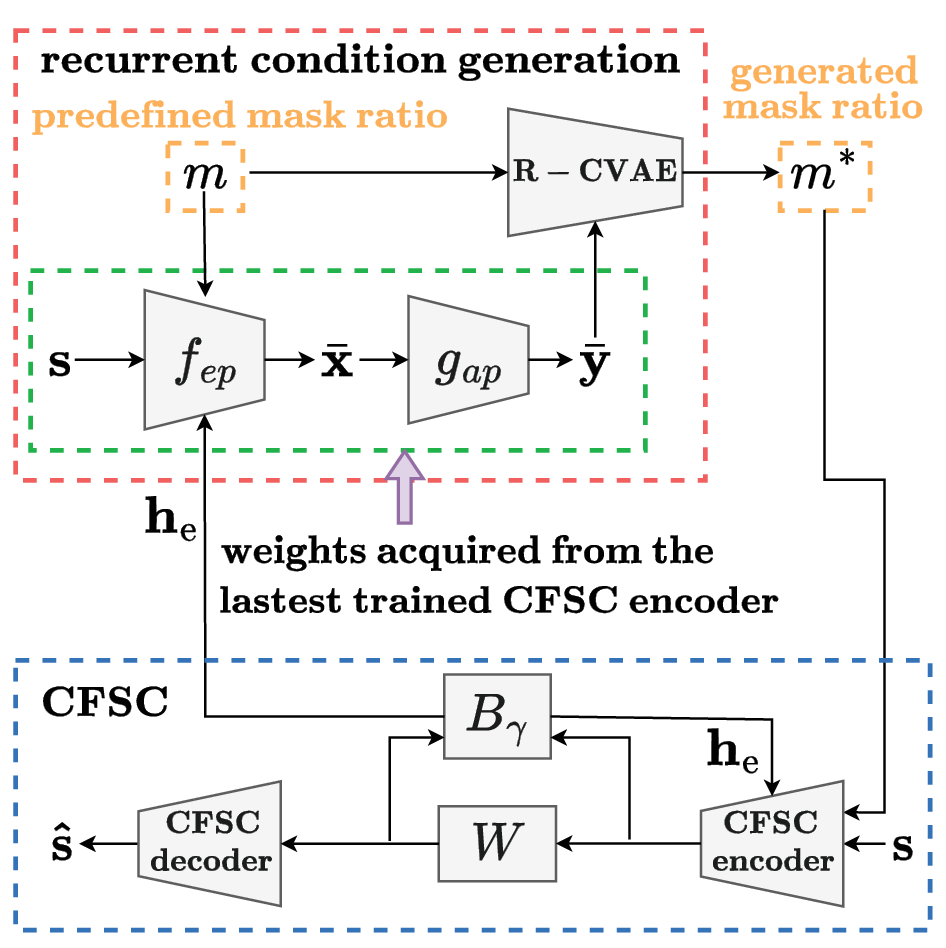}
	\caption{Learnable CSI fusion semantic communication framework.} 
	\label{fig_6}
\end{figure}

\subsection{The Recurrent Condition Generation Stage}

As mentioned in Section $\mathrm{\uppercase\expandafter{\romannumeral3}}$, CFSC conducts CSI fusion for masking feature elements with low semantic importance and subchannel quality. In this way, the mask ratio $m$ plays a key role for the semantic encoder, since it balances the weight between extracted semantics and MIMO CSI for the CSI fusion in NI-CFMA. A fixed $m$ brings the following risks to the CFSC. On one hand, an excessive $m$ leads to a catastrophic masking of most semantic elements, resulting in a drastic performance loss. On the other hand, a tiny $m$ means the encoder transmits almost all the features, which will inevitably causes resources waste. These insights motivate us to make $m$ learnable to adapt to various source and channel distributions. We further introduce the learnable mask ratio into the CFSC through a preprocessing recurrent condition generation stage and define the overall new structure as LCFSC framework.

The structure of the LCFSC is shown in Fig. \ref{fig_6}. It can be separated into two parts: recurrent condition generation and JSCC-based semantic transmission. The former is a new preprocessing stage to generate suitable $m$ while the latter is our original CFSC. Since $m$ needs to be updated instantly before the JSCC transmission, we define the proxy transmitted codewords, $\bar{\mathbf{y}}\in \mathbb{R}^{C_{\mathrm{L}}}$, as the proximate substitute of the actual transmitted $\mathbf{y}$, which are given by the CFSC encoder. The weights of proxy network are from the lastest trained CFSC encoder with the stable mask ratio $m$. The proxy transmitted $\bar{\mathbf{y}}$ can be expressed as 
\begin{align}
	\bar{\mathbf{y}}=f_p(\mathbf{s}, \mathbf{h}_{\mathrm{e}},m)=g_{ap}(f_{ep}(\mathbf{s}, \mathbf{h}_{\mathrm{e}},m)),
\end{align}
where $f_p(\cdot, \cdot, \cdot): \mathbb{R}^{H\times W\times C}\mapsto \mathbb{R}^{C_{\mathrm{L}}}$ refers to the proxy encoder network.

Then the stable $m$ and the proxy codewords, $\bar{\mathbf{y}}$, are fed into the CVAE-based condition generator \cite{VAE,CVAE} for the suitable $m^\ast$, which is shown as
\begin{align}
	m^\ast=\zeta(m, \bar{\mathbf{y}}),
\end{align}
where $\zeta(\cdot, \cdot)$ is the proposed recurrent conditional variational autoencoder (R-CVAE) network, $m^\ast$ is the output suitable mask ratio.

\subsection{R-CVAE for Generating the Suitable Conditions}

Obtaining the proper mask ratio for various image sources and channel conditions is an unsupervised learning problem. In other words, the proper mask ratio in current stage is unknown, which means that no standard mask ratio values exist for us to evaluate the learnable $m^\ast$ through direct criteria such as mean square error (MSE) or cross entropy (CE) loss. In this way, variational learning aided by the predefined condition $m$ seems to become a competitive method for generating $m^\ast$ through the learned latent representation, $\mathbf{z}$.

As mentioned above, we propose the R-CVAE to generate current proxy semantic codewords, $\bar{\mathbf{y}}$, with the condition $m$, which can be expressed as
\begin{align}
	p(\bar{\mathbf{y}}|m) & = \int p(\bar{\mathbf{y}},\mathbf{z}|m)d\mathbf{z} = \int p(\bar{\mathbf{y}}|\mathbf{z},m)p(\mathbf{z}|m)d\mathbf{z}.
\end{align}

However, the integral of the marginal likelihood $p(\bar{\mathbf{y}}|m)$ is intractable, hence variational interference is introduced to solve the problem. It aims to maximize the conditional log-likelihood, $\log_{}{p}(\bar{\mathbf{y}}|m)$. This term can be decoupled into $p(\bar{\mathbf{y}},\mathbf{z}|m)$ and $p(\mathbf{z}|\bar{\mathbf{y}},m)$, in which the latter is the true posterior of the latent representation. Since $p(\mathbf{z}|\bar{\mathbf{y}},m)$ is also intractable, we introduce $q(\mathbf{z}|\bar{\mathbf{y}},m)$ as an approximate substitute. The conditional log-likelihood can be written as
\begin{equation}
	\begin{aligned}
		\log_{}{p}(\bar{\mathbf{y}}|m) & = \mathbb{E}_{\mathbf{z}\sim q(\mathbf{z}|\bar{\mathbf{y}},m)}[\log_{}{p}(\bar{\mathbf{y}},\mathbf{z}|m)-\log_{}{p}(\mathbf{z}|\bar{\mathbf{y}},m)] \\
		& \overset{\text{(a)}}{=} D_{\mathrm{KL}}[q(\mathbf{z}|\bar{\mathbf{y}},m)||p(\mathbf{z}|\bar{\mathbf{y}},m)] \\ & + \mathbb{E}_{\mathbf{z}\sim q(\mathbf{z}|\bar{\mathbf{y}},m)} [\log_{}{p}(\bar{\mathbf{y}},\mathbf{z}|m)-\log_{}{q}(\mathbf{z}|\bar{\mathbf{y}},m)]\\  
		& \ge  \mathbb{E}_{\mathbf{z}\sim q(\mathbf{z}|\bar{\mathbf{y}},m)} [\log_{}{p}(\bar{\mathbf{y}},\mathbf{z}|m)-\log_{}{q}(\mathbf{z}|\bar{\mathbf{y}},m)],
	\end{aligned}
\end{equation}
where the first term $D_{\mathrm{KL}}[q(\mathbf{z}|\bar{\mathbf{y}},m)||p(\mathbf{z}| \bar{\mathbf{y}},m)]$ in $(\text{a})$ represents the differences between the true posterior and the approximation posterior distribution. However, the KL-Divergence \cite{KL} between these two distributions is hard to be computed. And since $\log_{}{p}(\bar{\mathbf{y}}|m)$ remains unchanged under certain image source and condition $m$, minimizing the KL-Divergence $D_{\mathrm{KL}}[q(\mathbf{z}|\bar{\mathbf{y}},m) ||p(\mathbf{z}| \bar{\mathbf{y}},m)]$ is equivalent to maximizing the second term in $(\text{a})$. In the common variational interference, the second term is named evidence lower bound (ELBO), which can be further rewritten as
\begin{equation}
\begin{aligned}
	\mathrm{ELBO} = & \mathbb{E}_{\mathbf{z}\sim q(\mathbf{z}|\bar{\mathbf{y}},m)}[\log_{}{p}(\bar{\mathbf{y}},\mathbf{z}|m)-\log_{}{q}(\mathbf{z}|\bar{\mathbf{y}},m)] \\ = & \mathbb{E}_{\mathbf{z}\sim q(\mathbf{z}|\bar{\mathbf{y}},m)}[\log_{}{p}(\bar{\mathbf{y}}|\mathbf{z},m)]\\ & -D_{\mathrm{KL}}[q(\mathbf{z}|\bar{\mathbf{y}},m)||p(\mathbf{z}|m)].
\end{aligned}
\end{equation}

From the last equation, we observe that the ELBO can be rewritten as the sum of two terms. The first term encapsulates the distortion, when reconstructed from the encoding $\mathbf{z}$ along with condition $m$. The second one is a regulation term that ensures the latent variables given $\bar{\mathbf{y}}$ and $m$ being close to the corresponding encoding given $m$.

In VAE, the latent variables $\mathbf{z}$ are sampled from a standard Gaussian distribution with reparametrization trick. We define that conditioned on $m$, the latent representation $\mathbf{z}$ is normally distributed with mean $f_{\mu}(\bar{\mathbf{y}})$ and a diagonal covariance matrix whose diagonal entries are given by ${\exp}({f_{\sigma}(\bar{\mathbf{y}})})$. Moreover, we approximate the posterior distribution of $\mathbf{z}$ given $\bar{\mathbf{y}}$ and $m$ by a normal Gaussian distribution with mean $h_{\mu}(\bar{\mathbf{y}},m)$ and a diagonal covariance matrix whose diagonal entries are given by ${\exp}({h_{\sigma}(\bar{\mathbf{y}},m)})$. In this way, the latent variables $\mathbf{z}$ can be given as
\begin{align}
	\mathbf{z}=h_{\mu}(\bar{\mathbf{y}},m)+\bm{\epsilon}\odot h_{\sigma}(\bar{\mathbf{y}},m),
\end{align}
where $\bm{\epsilon} \sim \mathcal{N}(0,\bm{I})$ denote the sampled normal Gaussian variables.

\begin{figure}[htbp]
	\centering
	\includegraphics[width=3.3in]{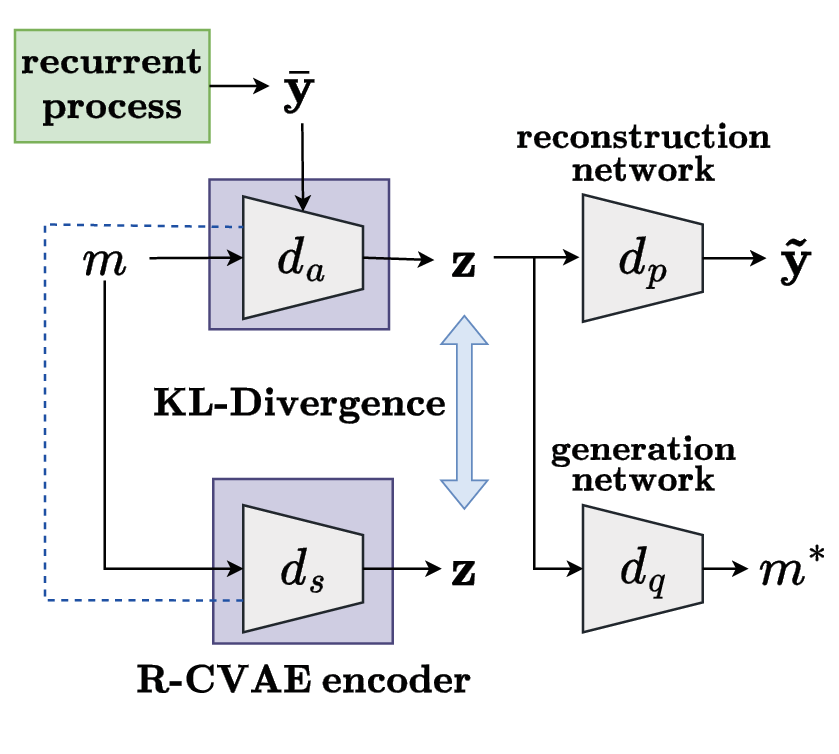}
	\caption{The structure of the proposed recurrent conditional variational autoencoder.}
	\label{fig_7}
\end{figure}

The structure of the R-CVAE is shown in Fig. \ref{fig_7}. $d_a(\cdot,\cdot)$ is the R-CVAE encoder representing the posterior distribution given $\bar{\mathbf{y}}$ joint with $m$, which generates the latent variables, $\mathbf{z}$. $d_s(\cdot)$ is the encoder with only $m$ input and regulates the latent variables generated from $d_a(\cdot,\cdot)$. Their KL-Divergence is computed as the second term in Eq. (17). Then the latent variables $\mathbf{z}$ pass through the R-CVAE decoder, $d_p(\cdot)$, to reconstructed the semantic codewords as $\mathbf{\tilde{y}}$. The MSE between $\bar{\mathbf{y}}$ and $\mathbf{\tilde{y}}$ refers to the first term in Eq. (17), since $p(\bar{\mathbf{y}}|\mathbf{z},m)$ follows the Guassian distribution. With the trained $\mathbf{z}$ from $d_a(\cdot,\cdot)$, we utilize such latent variables through a generation network, $d_q(\cdot)$, to receive the refined mask ratio $m^\ast$.

Before outputting the refined mask ratio, we predefine a vector $\bm{\bar{M}}=[m_{1}, m_{2}, \cdots, m_{K}]$, in which each element $m_{i}$ represents a specific mask ratio value. The recognition network outputs the weights of $\bm{\bar{M}}$ as $\bm{\beta}=[\beta_{1}, \beta_{2}, \cdots, \beta_{K}]$. Thus the mask ratio $m^\ast$ can be computed as 
\begin{align}
	m^\ast & = \bm{\bar{M}}\bm{\beta}^T.
\end{align}
\subsection{Training Loss}
Based on the analysis above, we then present the training loss function for the LCFSC step by step.

The CFSC framework adapts the typical JSCC structure. As a result, training loss is simply the direct reconstruction loss for images. In this paper, we denote $L_\mathrm{1}$ as the reconstruction loss, which can be expressed as
\begin{align}
	L_\mathrm{1}=\frac{1}{N}\sum_{j=1}^{N}D\left (\mathbf{\hat{s}}_j,\mathbf{s}_j  \right ),
\end{align}
where $N$ refers to the total number of source images, $\mathbf{s}_j$ refers to the $j$-th source image, $\mathbf{\hat{s}}_j$ refers to the $j$-th reconstructed image, $D(\cdot, \cdot)$ is the target distortion function between $\mathbf{s}_j$ and $\mathbf{\hat{s}}_j$.

For the LCFSC, the training loss depends both on the JSCC semantic transmission and the recurrent condition generation parts. Based on the rewritten ELBO in Eq. (17), the reconstruction loss and the regulation loss can be expressed as
\begin{align}
	L_\mathrm{rec} = \mathbb{E}_{\mathbf{z}\sim q(\mathbf{z}|\bar{\mathbf{y}},m)}[\log_{}{p}(\bar{\mathbf{y}}|\mathbf{z},m)]=\frac{1}{N}\sum_{j=1}^{N}||\tilde{\mathbf{y}}_j-{\bar{\mathbf{y}}}_j||^2,
\end{align}
\begin{equation}
	\begin{aligned}
		L_{\mathrm{reg}} = & D_{\mathrm{KL}}[q(\mathbf{z}|\bar{\mathbf{y}},m)||p(\mathbf{z}|m)] \\ = &  \frac{1}{2N}\sum_{j=1}^{N}\sum_{i=1}^{L}\bigg[f_{ji\sigma}(\bar{\mathbf{y}})-h_{ji\sigma}(\bar{\mathbf{y}},m) +\\ & \exp(h_{ji\sigma}(\bar{\mathbf{y}},m)-f_{ji\sigma}(\bar{\mathbf{y}}))+\\& \left.\frac{[f_{ji\mu}(\bar{\mathbf{y}})-h_{ji\mu}(\bar{\mathbf{y}},m)]^2}{\exp(f_{ji\sigma(\bar{\mathbf{y}})})} \right], 
	\end{aligned}
\end{equation}
where $\tilde{\mathbf{y}}_j$ represents the reconstructed transmitted codewords of the $j$-th image, ${\bar{\mathbf{y}}}_j$ represents the proxy transmitted codewords of the $j$-th image, $f_{ji\mu}$ and $h_{ji\mu}$ denote the $i$-th mean element of the function $f$ and $h$ with the $j$-th image while $f_{ji\sigma}$ and $h_{ji\sigma}$ are the $i$-th convariance matrix element of the function $f$ and $h$ with the $j$-th image, respectively. The sequence length of the latent representation is denoted as $L$.

To maximize the ELBO, the loss for the R-CVAE part can be expressed as
\begin{align}
	L_{\mathrm{c}} = L_{\mathrm{rec}} + L_{\mathrm{reg}}. 
\end{align}

Overall, combining the JSCC and R-CVAE part together, we get the training loss for the LCFSC, which can be formulated as
\begin{align}
	L_{\mathrm{2}} = L_{\mathrm{1}} + \lambda L_{\mathrm{c}}, 
\end{align}
where $\lambda$ is the trade-off term controlling both $L_1$ and $L_c$.

\subsection{Training Strategy}

\begin{algorithm}[htbp]
	\caption{Pretrain Method for the Noise Purified Channel Estimator}\label{alg:alg3}
	\begin{algorithmic}
		\STATE 
		$\textbf{Input:}$ Channel SNR, Pilot sequence $\bm{\tau}$, and Sample sets of MIMO CSI matrix $\mathbf{H}$		
		
		$\textbf{Output:}$ The noise purified channel estimator $B_{\bm{\Upsilon}}\left(\cdot, \cdot \right)$	
		
		\STATE \hspace{0.5cm}$ \textbf{} $
		
		\STATE \hspace{0.5cm}1. Take a batch $\mathbf{h}'$ from the sample sets $\mathbf{H}$
		\STATE \hspace{0.5cm}2. Let the stable pilot sequence $\bm{\tau}$ faces with generated 
		\STATE \hspace{0.9cm}MIMO fading channels:
		\STATE \hspace{0.9cm}$\hat{\bm{\tau}}=\mathbf{h}'\bm{\tau}+\mathbf{n}$
		\STATE \hspace{0.5cm}3. Adapt LS to obtain the coarse CSI: $\mathbf{h}_{\mathrm{LS}}=\hat{\bm{\tau}}\bm{\tau}^{-1}$
		\STATE \hspace{0.5cm}4. Use $B_{\bm{\Upsilon}}$ to get fine CSI: $B_{\bm{\Upsilon}}\left(\bm{\tau}, \bm{\hat{\tau}} \right)\longrightarrow \mathbf{h}_\mathrm{e}$
		\STATE \hspace{0.5cm}5. Compute the loss with MSE
		\STATE \hspace{0.5cm}6. Gradient descent update weight parameters

		\STATE \hspace{0.5cm} $ \textbf{} $

		\STATE\hspace{0.5cm} return $B_{\bm{\Upsilon}}\left(\cdot, \cdot \right)$		
	\end{algorithmic}
	\label{alg1}
\end{algorithm}

To fully exploit the potential of LCFSC, we divide the training strategy into two steps. First, NPN are pretrained for a reasonably satisfying estimated CSI aided by the sampled CSI data of various SNRs. The algorithm for pretraining the NPN is shown in alg. 1. Note that the channel state remains unchanged during the CSI feedback. After that, we jointly train the LCFSC together with NPN in an end-to-end manner. With the pretrained NPN as a suitable initial value, accurate feedback CSI are enabled in the learnable semantic communication framework, which helps the overall LCFSC converge fast and stable. 

\section{Numerical Results}
In this section, numerical results and analysis are presented to verify the effectiveness of LCFSC compared to existing semantic communication and traditional separated source-channel coding (SSCC) schemes. 

\subsection{Experimental Setups}

\subsubsection{Datasets}

For the wireless semantic image transmission, we quantify the performances of LCFSC versus other benchmarks over the UDIS-D \cite{UDIS-D} dataset. It is a comprehensive real-world dataset, which contains a variety of common scenes. It has over 10,000 images for training and over 1,000 ones for testing, where images are shot under different perspectives and illumination conditions. During model training, images are resized into the shape of 128$\times$128$\times$3.

\subsubsection{Model Deployment Details}

With the aforementioned network details, we utilize the Swin Transformer backbone as the semantic codec and the Channel ModNet \cite{WITT} as the channel codec. $f_{e}$ and $f_{d}$ share the same network deployment with $\{N_1,N_2,N_3,N_4\}=\{2,2,6,2\}$ Transformer blocks. To present LCFSC in a more convenient way, the NI-CFMA modules embedded in $f_{e}$ only exist in the third stage of the $N_3$ Transformer blocks. In order to ensure proper mask ratio acquisition, we set the range of $m$ from 0.001 to 0.015, as a smaller mask ratio tends to retain the essential semantics while discarding unnecessary and noise-related elements. In this way, the pre-set mask ratio vector is defined as $\bm{\bar{M}} = [0.001, 0.003, 0.005, 0.007, 0.009, 0.012, 0.014, 0.015]$. For model training, we use variable learning rate, which decreases step-by-step from 1e-4 to 2e-5. The batchsize is set as 16. The whole framework is optimized with Adam \cite{Adam}. All the experiments of LCFSC and other DL-based benchmarks are runned in RTX3090 GPUs. Unless specifically marked, we choose multi-scale structural similarity (MS-SSIM) \cite{ssim} for the basic reconstruction loss $L_\mathrm{1}$ while setting MIMO antenna numbers as $N_\mathrm{T} = 2$ and $N_\mathrm{R} = 2$. To simulate the practical 5G MIMO transmission scenarios, MIMO CSI matrices are generated according to \cite{channel}, a general non-stationary 5G wireless channel model. 1000 samples of MIMO channel matrices are utilized for image transmission during training while 100 extra samples are ready for testing, ensuring the performance evaluation over practical channel states.

\subsubsection{Comparison Benchmarks}
In the experiments, several benchmarks are given as below

\begin{itemize}
	\item[$\bullet$] NTSCC: The nonlinear transform source-channel coding \cite{NTSCC} with the Swin Transformer backbone, generating extra rate tokens as side information into the Transformer structure through the concatenation of latent variables and MIMO CSI matrix.
	\item[$\bullet$] ADJSCC: The adaptive deep JSCC scheme in \cite{ADJSCC} with the convolution neural network (CNN) structure, directly blending both MIMO CSI and SNR as side information with original features in attention modules.
	\item[$\bullet$] FLSC with single user: The JSCC-guided semantic communication framework \cite{flsc} with the proposed hierarchical Vision Transformer as network backbone.	
	\item[$\bullet$] JPEG+LDPC+QAM+SVD: The traditional coding transmission scheme with the joint photographic experts group (JPEG) as the source coding and the low density parity check (LDPC) code as the channel coding scheme, enhanced by the singular value decomposition (SVD) for the common MIMO precoding and detection. Quadrature amplitude modulation (QAM) is utilized as the modulation scheme.
	\item[$\bullet$] BPG+Capacity: The traditional coding transmission scheme with powerful BPG codec \cite{bpg} for compression while applying ideal channel capacity achieving code.
\end{itemize}

\begin{figure*}[htbp]
	\centering  %图片全局居中
	\subfigure[PNSR for the reconstructed images.]{
		\includegraphics[width=0.49\linewidth]{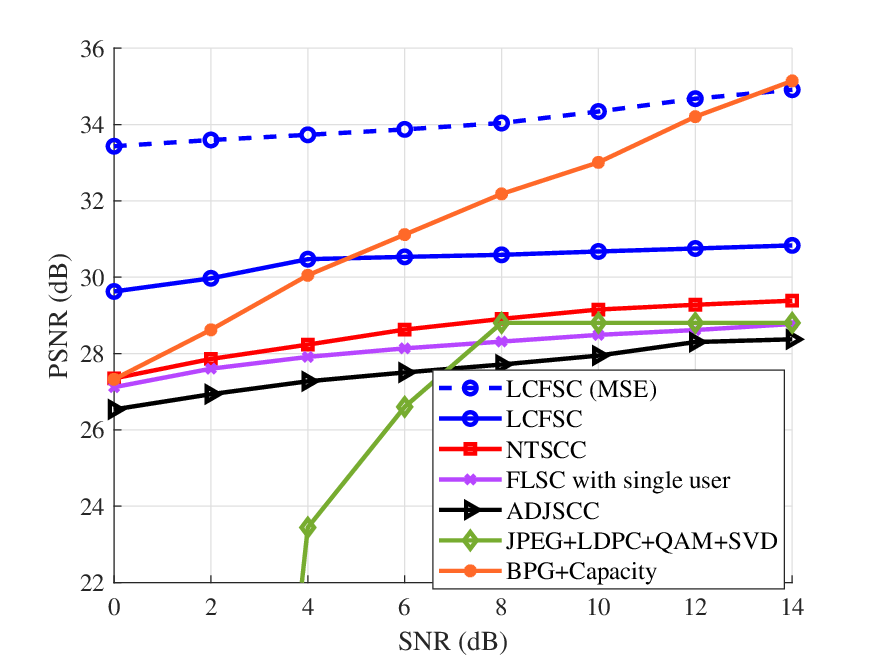}}
	\subfigure[MS-SSIM for the reconstructed images.]{
		\includegraphics[width=0.49\linewidth]{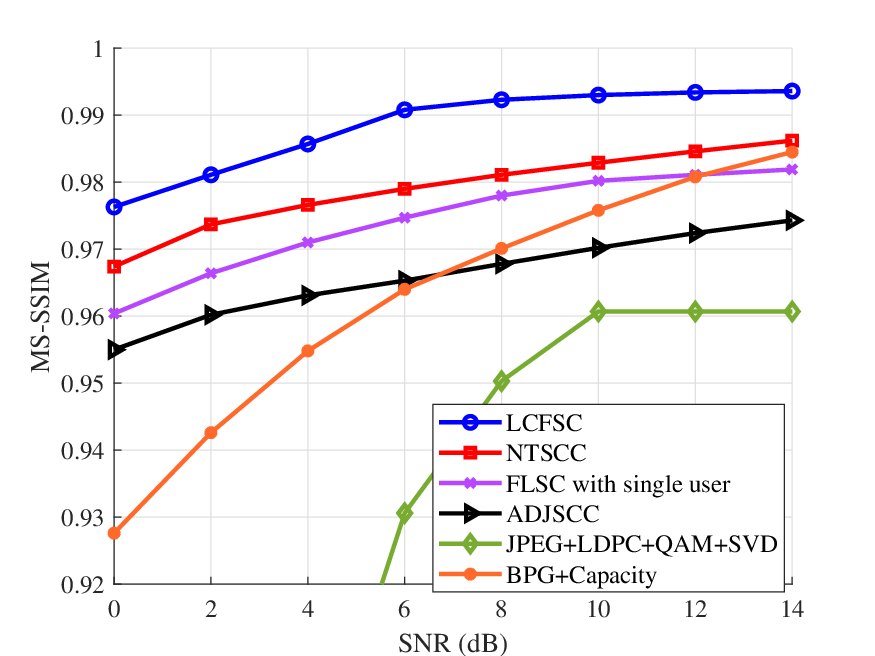}}
	\caption{Quality of the reconstructed images versus the SNRs in MIMO fading channels ($R$ = 0.06).}
	\label{fig_8}
\end{figure*}

\begin{figure*}[htbp]
	\centering  %图片全局居中
	\subfigure[PNSR for the reconstructed images.]{
		\includegraphics[width=0.49\linewidth]{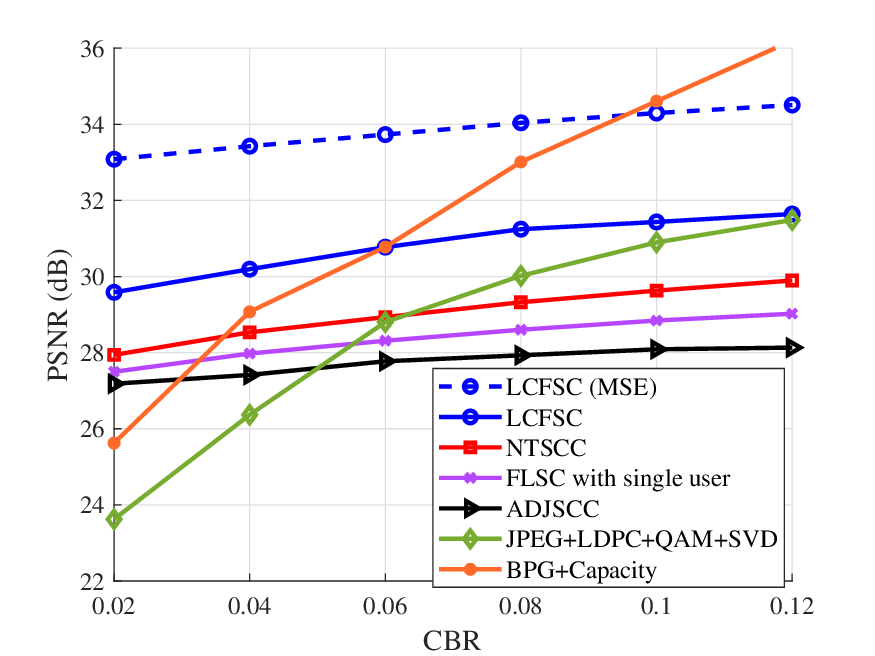}}
	\subfigure[MS-SSIM for the reconstructed images.]{
		\includegraphics[width=0.49\linewidth]{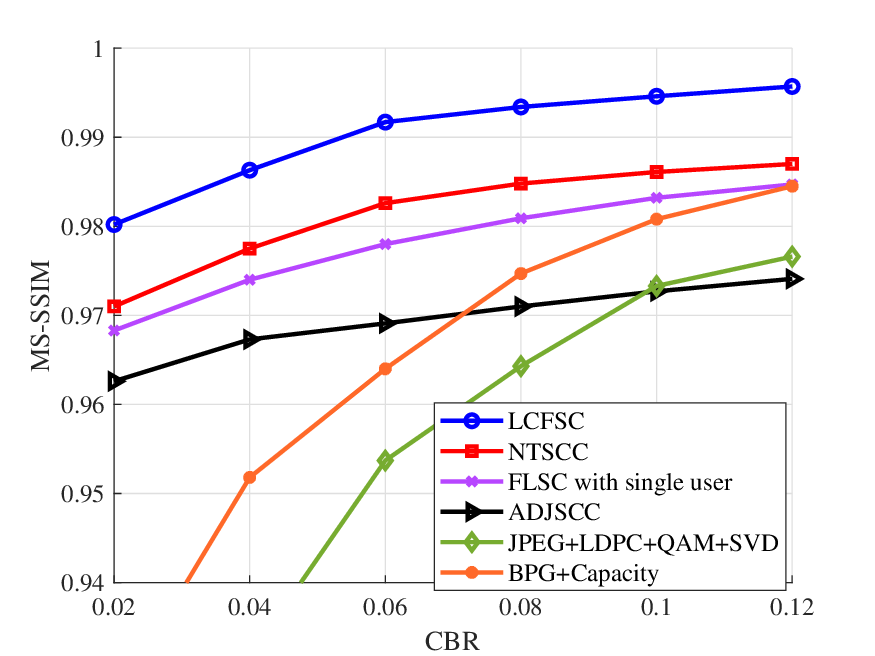}}
	\caption{Quality of the reconstructed images versus the CBRs in MIMO fading channels (SNR = 8 dB).}
	\label{fig_9}
\end{figure*}

For the NTSCC, it is an up-to-date JSCC-based semantic communication framework with the Swin Transformer backbone aided by the side information including latent hyperprior and MIMO CSI for conducting the rate allocation, which is corresponding to the scheme in Fig. 1(a). For the ADJSCC, it is assumed as the representative work for intuitively incorporating the feedback CSI as side information into the deep JSCC design, which is corresponding to the scheme in Fig. 1(b). Both NTSCC and ADJSCC combine CSI with semantic features through intuitive concatenation. For the FLSC, we consider the single-user scenario and mainly focus on the different transformer structure as network backbone. For the JPEG+LDPC+QAM+SVD, we assume it as a totally traditional SSCC scheme in practice. For the BPG+Capacity, it is an ideal scheme for the traditional SSCC.

\subsubsection{Evaluation Metrics}

We leverage the widely used pixel-wise metric peak signal-to-noise ratio (PSNR) and the perceptual-level MS-SSIM as measurements for the reconstructed image quality. 

\subsection{Results Analysis}

\subsubsection{SNR Performances}

We first present the SNR performances for the LCFSC and other benchmarks in Fig. \ref{fig_8}. From Fig. \ref{fig_8}(a), It is clearly to observe that LCFSC outperforms NTSCC in all ranges of SNRs and the performance gap even increases when the SNR value drops. Especially in low SNR such as 0 dB, LCFSC surpasses NTSCC over 2 dB in terms of PSNR. It is because that NTSCC mainly utilizes nonlinear transform coding for adaptively adjusting the transmission rate, which is designed to fit both the source and channel distribution to acquire the performance gain. However, when faced with MIMO fading channels, even aware of MIMO CSI, it brings much more difficulty for the NTSCC to adapt the channel distribution under specific CBRs. As such, it also brings extra problems for the source codec to better extract and translate semantics. Due to the CSI-aware ability of LCFSC inside the semantic extractor, the burden for the channel coding and semantic understanding can be alleviated to some extent. For the ADJSCC, such direct concatenation operation for MIMO CSI fusion seems to be invasive to the semantic codec. LCFSC even obtains much more performance gain compared to ADJSCC than NTSCC. In this way, the advantage of incorporating CSI into semantic encoder over common schemes in Fig. 1(a)(b) can be verified. For the FLSC with single user, it performs better than ADJSCC. Since ADJSCC is a CNN-based framework, the superiority of Vision Transformer backbone has been verified. Built from Swin Transformer, NTSCC achieves better results than FLSC constructed by the hierarchical Vision Transformer. It further stresses the merit of Swin Transformer-based semantic communication frameworks. When compared to the traditional separated coding schemes, LCFSC is free from the cliff effect, presenting robust performances in face of various SNRs than the practical SSCC scheme JPEG+LDPC+QAM+SVD. Moreover, compared to the ideal separated coding scheme BPG+Capacity, LCFSC still performs better in a large range of SNRs when adopting MSE as reconstruction loss $L_\mathrm{1}$. In Fig. \ref{fig_8}(b), LCFSC outperforms all other benchmarks including DL-based schemes and traditional SSCC schemes for MS-SSIM, which also demonstrates that LCFSC not only reconstructs images well but also ensures good visual quality for human perception under MIMO fading channels with different noise intensity levels.

\subsubsection{CBR Performances}

Then we evaluate the CBR performances in Fig. \ref{fig_9}. CBR represents the compression ratio between the transmitted sequence length and original image size. In general, the LCFSC generally outperforms other DL-based schemes in all CBRs for both PSNR and MS-SSIM metrics in MIMO transmission scenarios. Even in extreme low CBR such as 0.02 or 0.04, LCFSC still achieves relatively satisfying performances, which indicates the superiority of utilizing semantics for the efficient data compression and transmission. When it comes to traditional separated coding schemes, the quality of reconstructed images decreases rapidly with the drop of CBRs. Since LCFSC enables adaptively adjusting the source and channel coding rate based on deep JSCC structure while ensuring the CSI-aware performances through robust semantic coding, it performs to be efficient in different channel bandwidth conditions.

\begin{figure*}[htbp]
	\centering
	\includegraphics[width=7.0in]{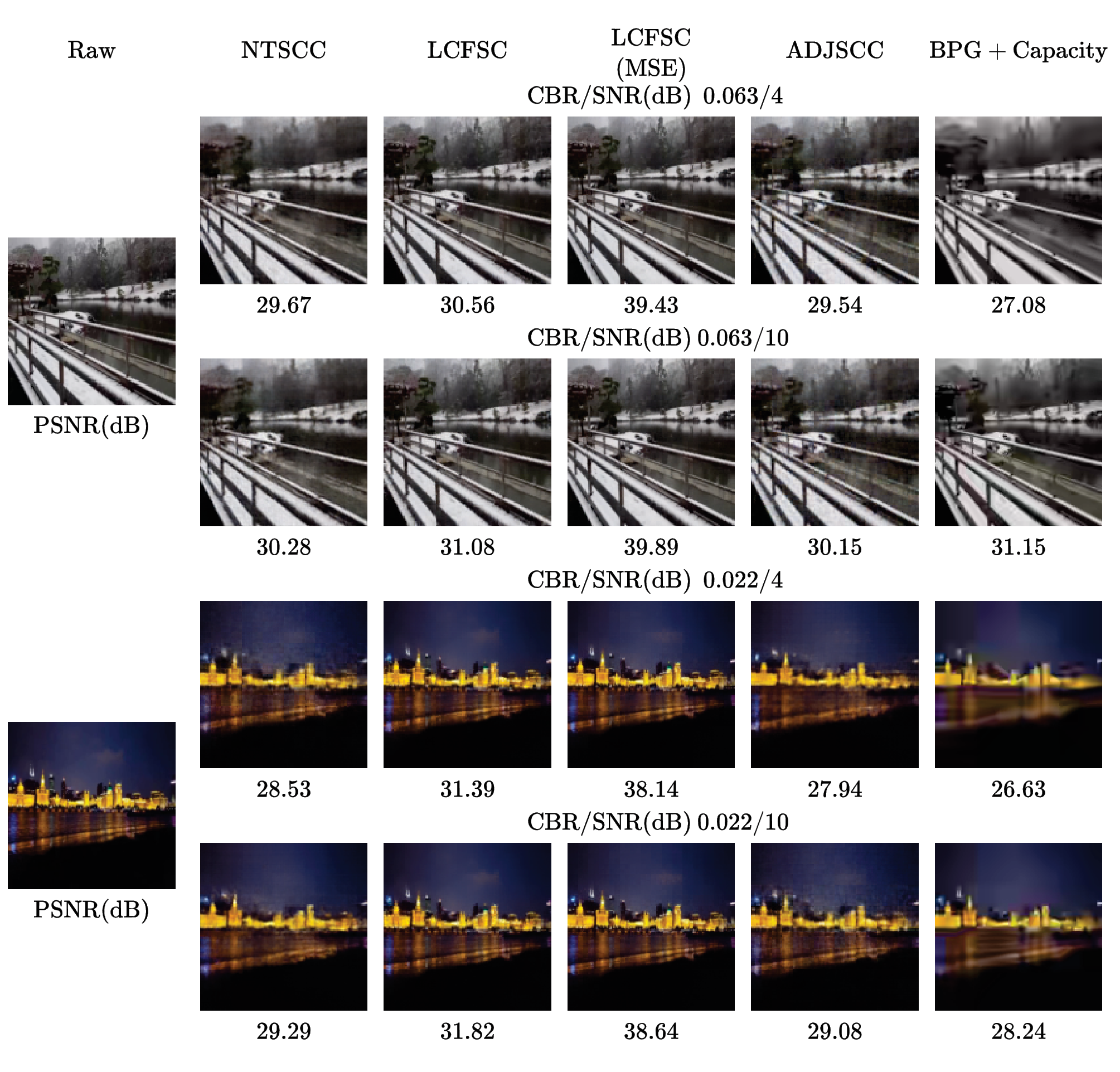}
	\caption{Examples of visual results. The first to the fifth row presents different CBR and SNR conditions for LCFSC and other comparable benchmarks.}
	\label{fig_10}
\end{figure*}

\subsubsection{Visual Results}

As shown in Fig. \ref{fig_10}, reconstructed images of LCFSC and other comparable benchmarks are presented under various CBR and SNR conditions. Compared to those DL-based semantic communication frameworks, LCFSC is capable of achieving higher visual quality under the same CBR or SNR condition. In other words, more details of the original images are retained compared to NTSCC and ADJSCC, which is especially friendly to the human vision. When it comes to the traditional separated coding scheme, LCFSC obviously outperforms BPG+Capacity under low CBR scenarios such as 0.022 or 0.063. Moreover, when we apply MSE as the basic reconstruction loss $L_\mathrm{1}$, the evaluated PSNR increases a lot, providing much clearer reconstructed images compared to other schemes. The visual results further intuitively demonstrate the stability of the LCFSC under MIMO fading channels.

\subsubsection{Simulations for different MIMO modes}

\begin{table}[htbp]
	\centering
	\caption{Evaluation of different MIMO modes. Red marks refer to the performance gap between the framework with and without the CSI fusion masking attention module.}
	\label{table1}
	
	\begin{tabular}{|c|c|c|}  
		\hline 
		& &\\[-6pt] 
		MIMO type&PSNR (dB)& $\bar{m}$ \\
		\hline
		& &\\[-6pt]  
		2$\times$2 diversity/multiplexing&32.677 ({\color{red}{+0.713}})&0.0062 \\
		\hline
		& &\\[-6pt]  
		4$\times$4 diversity/multiplexing&31.483 ({\color{red}{+0.742}})&0.0064 \\
		\hline
		& &\\[-6pt]  
		8$\times$8 diversity/multiplexing&30.715 ({\color{red}{+0.783}})&0.0079 \\
		\hline
		& &\\[-6pt] 
		16$\times$16 diversity/multiplexing&30.096 ({\color{red}{+0.867}})&0.0088 \\
		\hline
	
	\end{tabular}
\end{table}

We then test the performance of LCFSC in different MIMO modes. For the MIMO modes in semantic communication, diversity and multiplexing can be exploited, respectively. Diversity is achieved by the multiple antennas at the receiving end, as the same data stream is able to be transmitted and received on independent wireless links, greatly alleviating MIMO fading brought by wireless channels. Multiplexing is achieved by the multiple antennas at the transmitting end, making different semantics transmit in different streams simultaneously. We provide simulated results for MIMO diversity/multiplexing modes with various antenna numbers. To ensure the equal comparison, we generate different MIMO fading channel data with the same parameters except the transceiver antenna numbers. The feedback CSI is assumed as perfect to the transmitting end as well. From the Tab. \ref{table1}, the obvious trend can be seen that more antennas seem to bring worse performances than less antennas. Larger antenna array scale means that more interference among different antennas exists, which is more difficult for LCFSC to generate robust transmitted codewords adjusting to the channel states. However, as shown in the data marked in red, more severe interference among different antennas brings bigger performance gap for the Swin Transformer-based framework with and without CSI fusion masking attention module. It is because that the core performance gain of LCFSC is the non-invasive utilization of MIMO CSI, which exists in the form of the CSI fusion masking mechanism inside the attention module. Moreover, more antennas present the trend of larger average mask ratio $\bar{m}$, which means that the network has to mask more semantic elements to combat MIMO fading brought by a larger antenna scale.

\subsection{Ablation Study for the Proposed Modules}

\begin{figure}[htbp]
	\centering
	\includegraphics[width=3.6in]{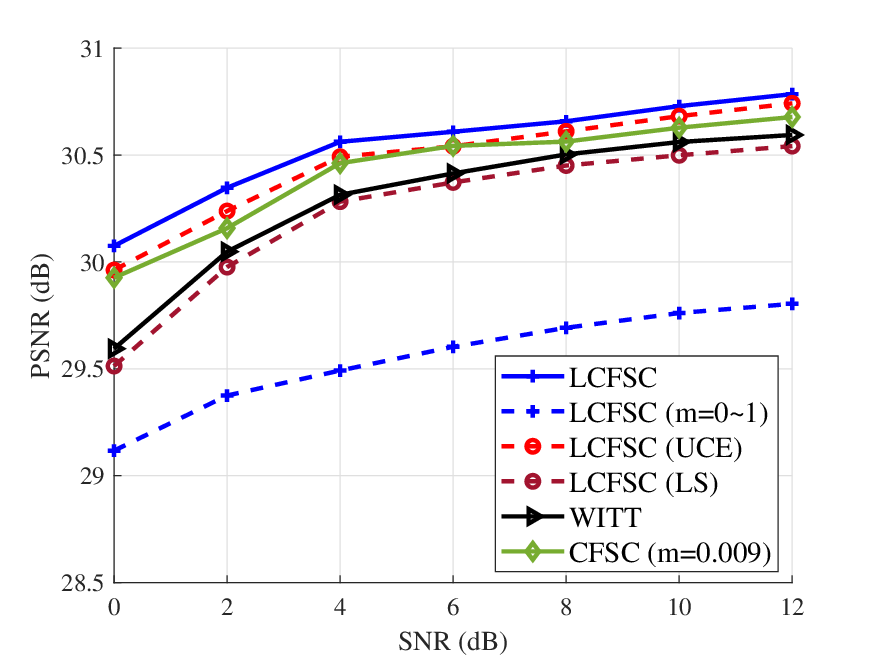}
	\caption{Abalation study for the CSI fusion masking module and NPN channel estimator. (CBR = 0.06, 8$\times$8 MIMO)}
	\label{fig_11}
\end{figure}

To verify the effectiveness of proposed LCFSC framework, we present several extra competitors in Fig. \ref{fig_11}. One of them is the wireless image transmission transformer (WITT) \cite{WITT}, which is a JSCC-based semantic communication framework with Swin Transformer as the semantic codec and Channel ModNet as the channel codec. Our proposed CFSC is presented as well, which has different stable mask ratios. Then LCFSC with the learnable mask ratio ranging from 0 to 1 is named as LCFSC ($m$=0$\sim$1). LCFSC with different channel estimation methods, e.g. U-channel estimator \cite{flsc} and traditional LS channel estimator are given as LCFSC (UCE) and LCFSC (LS), respectively. All of these schemes share the same structure and configuration as LCFSC for the same network components. 

We can easily find the trend that CFSC ($m$ = 0.009) exceeds the WITT, illustrating the performance gain brought by CSI fusion masking for robust semantic coding. Moreover, the result of LCFSC ($m$=0$\sim$1) is worse than the CFSC ($m$ = 0.009), even inferior to WITT. This phenomenon certificates that mask ratio is a crucial hyperparameter in the CFSC and improper $m$ value causes invasive effect, which means that the side information CSI overwhelms the raw semantic features. The larger $m$ value is, more semantic features are masked, which may inevitably discard some key semantics or those semantic elements faced with mild channel fadings. As such, a performance limitation for LCFSC ($m$=0$\sim$1) is emerged, as PSNRs increase extremely slow in high SNRs. Actually, with proper pre-set mask ratio vector $\bm{\bar{M}}$, the proposed LCFSC can easily solve this intractable problem through making the mask ratio learnable, adaptively matching the source and channel conditions. In this way, LCFSC outperforms both WITT and CFSC explicitly. When it comes to the accurate MIMO CSI acquisition in the whole semantic communication framework, LCFSC with the proposed NPN also outperforms LCFSC (UCE) and LCFSC (LS). It lies in the reason that the two-step channel estimation method is friendly to acquire relatively accurate CSI in a whole semantic network while the designed NPN structure is more efficient to purified estimated error brought by coarse LS estimation. In this way, our CSI fusion masking module and NPN are verified effective during the wireless MIMO image transmission.

\subsection{Analysis of Learnable Mask Ratio and CSI Fusion Attention Map}

To provide detailed analysis and interpretation to the effectiveness of proposed non-invasive CSI fusion masking attention module, the learnable mask ratio and attention mask maps are studied.

\subsubsection{Trend for the Learnable Mask Ratio}

\begin{figure}[htbp]
	\centering
	\includegraphics[width=3.6in]{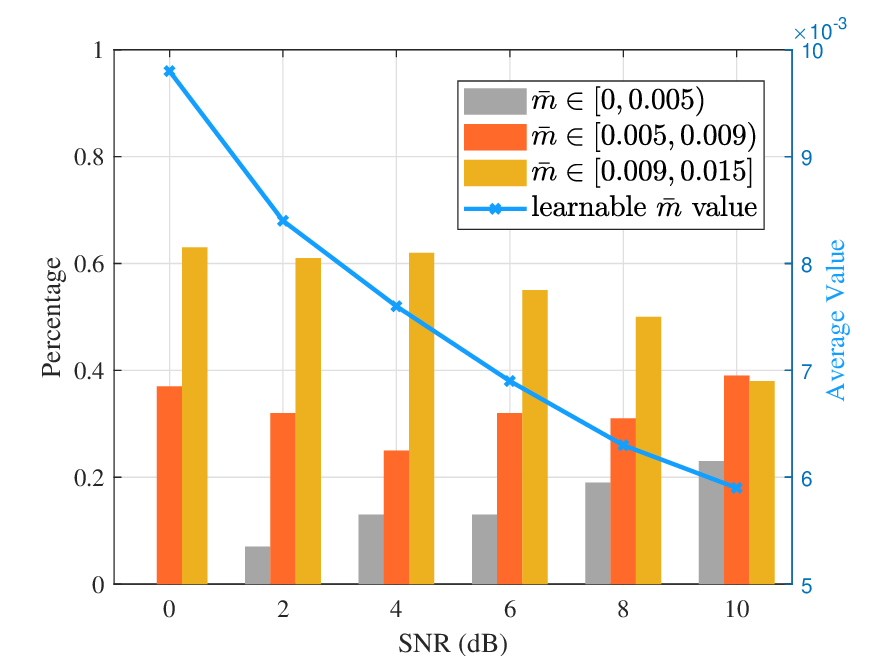}
	\caption{The changing values and distribution of learnable mask ratios.}
	\label{fig_12}
\end{figure}

We evaluate the changing of learnable mask ratio $\bar{m}$ along with the mask ratio distribution in different SNR conditions. From Fig. \ref{fig_12}, the marked broken line represents the average mask ratio value changing with SNRs while histograms show the percentage of $\bar{m}$ value distribution in different data ranges. For the changes of mask ratio value, we can explicitly observe that $\bar{m}$ seems to decrease with the increase of SNR values. This trend is in accordance with the target of proposed CSI fusion masking module as those elements with less semantic importance or faced with serious noise interference are to be discarded. Higher SNR condition represents a friendly channel state for MIMO image transmission, in which less semantic elements are interfered by channel fading and noise. As such, lower $\bar{m}$ is necessary for NI-CFMA to retain more necessary semantics. Vice versa, when the SNR is low, the number of those relatively less important but highly noise-related elements can increase a lot, requiring a high mask ratio value. When it comes to the distribution of mask ratio value, the moderated range $\bar{m}\in [0.005, 0.009)$ performs relatively stable while the low range $\bar{m}\in [0, 0.005)$ and the high range $\bar{m}\in [0.009, 0.015]$ vary according to different SNRs. The changing trend of $\bar{m}$ and its distribution further demonstrate the effectiveness of proposed learnable mask ratio strategy.

\subsubsection{Interpretability of the Learned Attention Masking Map}

\begin{figure*}[htbp]
	\centering
	\includegraphics[width=6.9in]{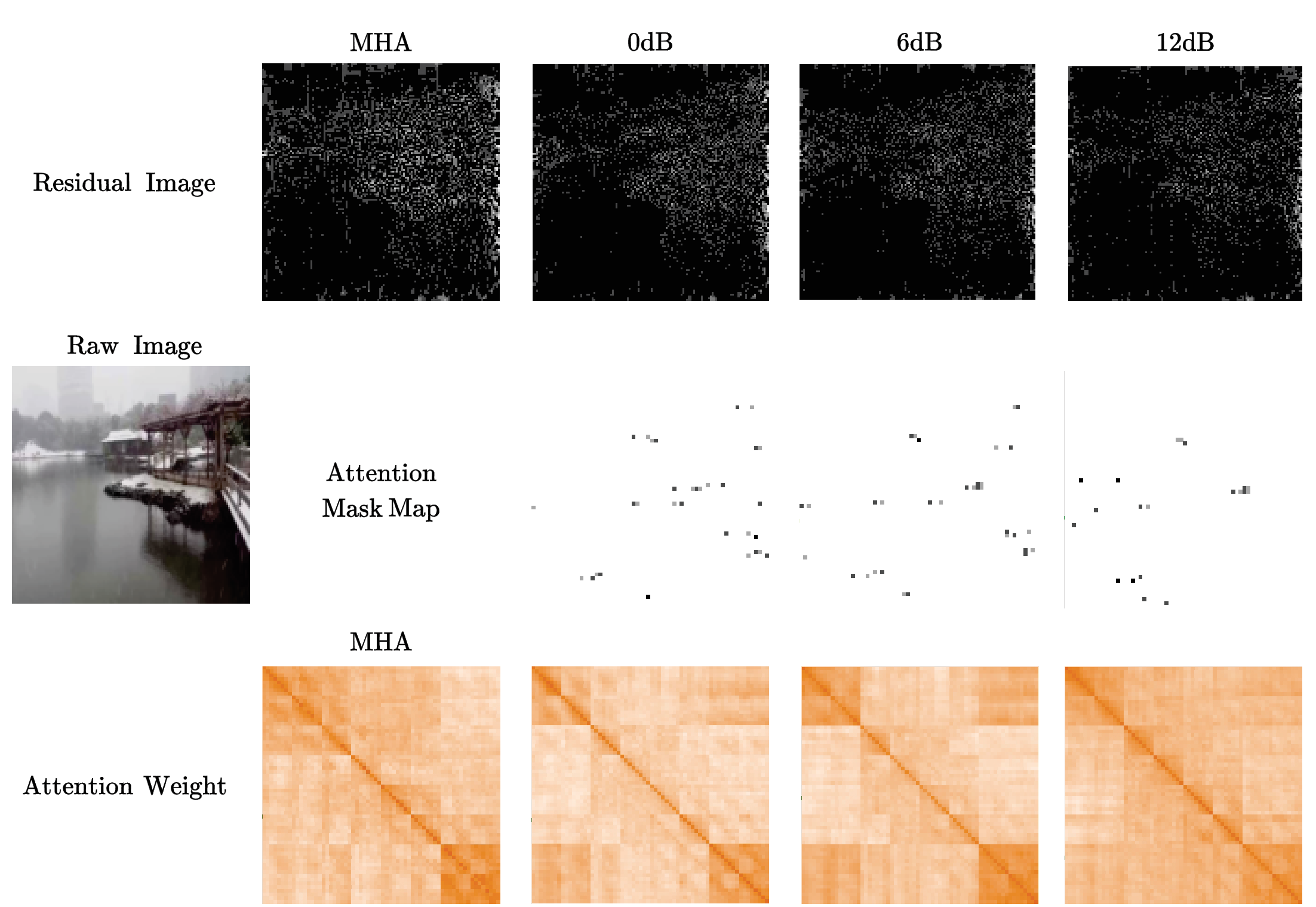}
	\caption{Interpretation of the CSI fusion attention map, including residual images, attention mask maps, and attention weights in different SNRs.}
	\label{fig_13}
\end{figure*}

\begin{figure}[htbp]
	\centering
	\includegraphics[width=3.6in]{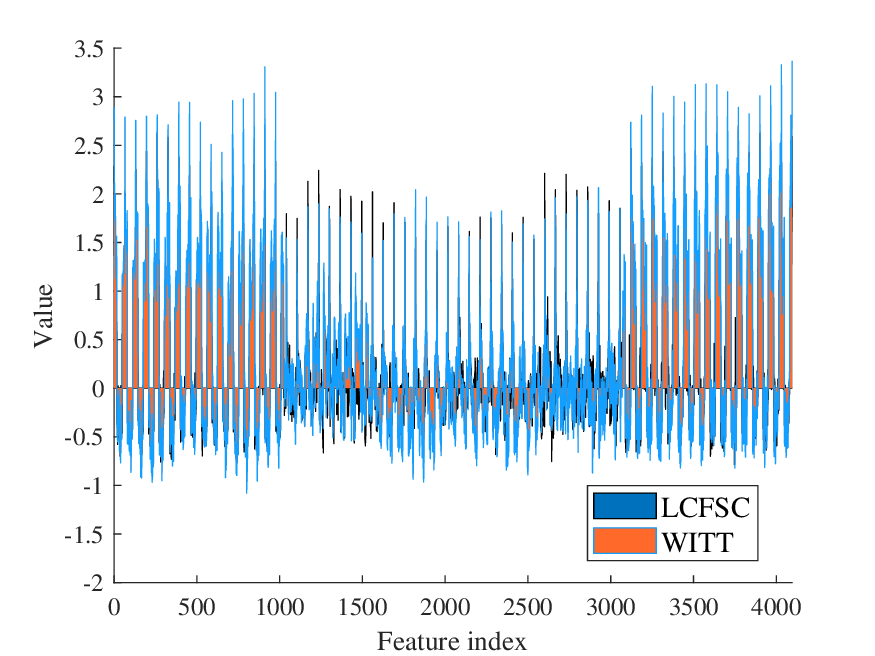}
	\caption{The distribution of attention weight values.}
	\label{fig_14}
\end{figure}

To give more interpretability for the performance gain of our proposed LCFSC, we further provide explanation for the CSI fusion masking mechanism in the attention module. As shown in Fig. \ref{fig_13}, residual images, along with attention mask maps and attention weight maps are shown in different SNR conditions. Residual image refers to the differences between the orginal image and reconstructed image, which presents the distribution of reconstruction error. With the original image and residual images, we can clearly observe that the reconstruction error mainly focuses on the foreground, such as garret and mountain, rather than the background lake and sky. With low SNRs, this phenomenon even becomes more obvious. It is because that foreground has much more complicated structure, edge, and color variation than the background, intrinsically harder to be reconstructed. However, when it comes to the scheme adopting common MHA, the error for the foreground is much more serious than those with NI-CFMA, which verifies the effectiveness of proposed NI-CFMA for outputting robust semantic codewords. After that, the attention mask maps are presented, where the masks are also mainly concentrated on the foreground part. In line with previous experiments, as the SNR decreases, more elements are to be masked to combat MIMO fading. Finally, the attention weight maps are also given. Compared to the attention weight distribution without CSI fusion masking strategy, the regenerated attention weight distributions show a stronger pattern in terms of locality, presenting more concentration on the diagonal. Then we further testify the effectiveness of CSI fusion masking strategy in terms of sharpening the distribution of attention weight. As shown in Fig. \ref{fig_14}, the attention weight distribution of LCFSC is similar to the one of WITT. However, the distribution is much sharper, reflecting the semantic enhancement effect. The above observations demonstrate that the NI-CFMA, which takes MIMO CSI as an auxiliary for calculating attention weight matrix, is able to learn robust attention weight distribution and hence improve the transmission performance.

\subsection{Analysis of Model Complexity and Computation Cost}

\begin{table}[htbp]
	\centering
	\caption{Evaluation of complexity and computation cost.}
	\label{table2}
	
	\begin{tabular}{|c|c|c|c|c|}  
		\hline 
		& & & &\\[-6pt] 
		Metric&LCFSC&NTSCC&ADJSCC&SSCC \\
		\hline
		& & & &\\[-6pt]  
		FLOPs (G)&15.2&16.9&3.3&/ \\
		\hline
		& & & &\\[-6pt]  
		Throughputs&161.4&147.9&327.3&6.5 \\
		\hline
		& & & &\\[-6pt] 
		Parameters (M)&58.4&53.8&60.8&/ \\
		\hline
	\end{tabular}
\end{table}

Finally, to evaluate the feasibility of LCFSC in practical deployment, we analyse the complexity and computation cost of proposed LCFSC and other benchmarks. Three metrics are selected to conduct this test. FLOPs represents the number of floating-point operations that can be performed per second, which measures the computational complexity of each model. Throughputs means the inference speed for trasmitting images, which refers to the number of images a model can transmit per second. Parameters is the total parameters of a model, which is mainly related to framework complexity. As shown in Tab. \ref{table2}, with similar Parameters, LCFSC shows competitive performance compared to NTSCC in terms of FLOPs and Throughputs. While for ADJSCC, since it is a thorough CNN-based framework, the computational cost can be much lower and inference speed can be much faster than the Swin Transformer-based schemes. It turns to be a trade-off between framework complexity and transmission accuracy. When it comes to traditional SSCC, the inference speed is much slower than DL-based schemes. When faced with poor channel conditions, traditional channel codec, e.g. LDPC, may consume much more time to reconstruct compressed images. When fully being trained and deployed in practice, LCFSC achieves satisfying throughputs and resonable model complexity. Practical techniques such as paralell computing and GPU acceleration can further optimize the deployment of LCFSC in practical scenarios.

\section{Conclusion}
This paper has proposed a novel framework which conducts robust semantic coding through incorporating feedback CSI into the semantic extractor with Swin Transformer backbone under practical general 5G MIMO image transmission scenario. The inner mechanism of attention module in the Transformer is explored through proper attention masking map which is produced by semantic features and the feedback CSI. The percentage of masking elements is controlled by the mask ratio during the recurrent condition generation stage, which mainly relies on the recurrent conditional variational autoencoder. Experiment results show the satisfying performances of LCFSC over traditional SSCC schemes such as JPEG+LDPC+QAM+SVD and BPG+Capacity along with the up-to-date NTSCC semantic communication schemes, which indicates the successful attempt of integrating MIMO CSI as side information into the semantic extractor for robust semantic coding. In LCFSC, the Swin Transformer-based semantic extractor can be aware of the current channel state and adaptively mask unimportant attention element for higher-quality wireless transmission. Further channel or rate adaptation designs can potentially enhance this structure in the future.

\fontsize{8pt}{10pt}\selectfont
%\bibliographystyle{IEEEtran}
%\bibliography{Ref}

\vfill

\end{document}